\documentclass[preprintnumbers,unsortedaddress,superscriptaddress,notitlepage]{revtex4-1}

\usepackage{amsfonts}
\usepackage{amsmath}
\usepackage[colorlinks=true,urlcolor=magenta,citecolor=magenta]{hyperref}
\usepackage{amssymb}
\usepackage{multirow}
\usepackage[english]{babel}
\usepackage[utf8]{inputenc}
\usepackage{graphicx}
\usepackage{epsfig}
\usepackage{bm}
\usepackage{longtable}
\usepackage{verbatim}
\usepackage{longtable}
\usepackage{booktabs}
\usepackage{color}
\usepackage{subfigure}
\usepackage{float}
\usepackage{cancel}

\newcommand{\mathsym}[1]{{}}

\newcommand{\emp}{\begin{equation}}
\newcommand{\fin}{\end{equation}}
\newcommand{\empn}{\begin{equation*}}
\newcommand{\finn}{\end{equation*}}

\newcommand{\bea}{\begin{eqnarray}}
\newcommand{\eea}{\end{eqnarray}}
\newcommand{\eger}{\begin{gather}}
\newcommand{\fger}{\end{gather}}
\newcommand{\egn}{\begin{gather*}}
\newcommand{\fgn}{\end{gather*}}
\newcommand{\bit}{\begin{itemize}}
\newcommand{\eit}{\end{itemize}}

\usepackage{tikz}
\usetikzlibrary{arrows,shapes}
\usetikzlibrary{trees}
\usetikzlibrary{matrix,arrows} 				
\usetikzlibrary{positioning}				
\usetikzlibrary{calc,through}				
\usetikzlibrary{decorations.pathreplacing}  
\usepackage{pgffor}							

\usetikzlibrary{decorations.pathmorphing}	
\usetikzlibrary{decorations.markings}
\tikzset{
    vector/.style={decorate, decoration={snake}, draw},
	provector/.style={decorate, decoration={snake,amplitude=2.5pt}, draw},
	antivector/.style={decorate, decoration={snake,amplitude=-2.5pt}, draw},
    fermion/.style={draw=black, postaction={decorate},
        decoration={markings,mark=at position .55 with {\arrow[draw=black]{>}}}},
    fermionbar/.style={draw=black, postaction={decorate},
        decoration={markings,mark=at position .55 with {\arrow[draw=black]{<}}}},
    fermionnoarrow/.style={draw=black},
    gluon/.style={decorate, draw=black,
        decoration={coil,amplitude=4pt, segment length=5pt}},
    scalar/.style={dashed,draw=black, postaction={decorate},
        decoration={markings,mark=at position .55 with {\arrow[draw=black]{>}}}},
    scalarbar/.style={dashed,draw=black, postaction={decorate},
        decoration={markings,mark=at position .55 with {\arrow[draw=black]{<}}}},
    scalarnoarrow/.style={dashed,draw=black},
    electron/.style={draw=black, postaction={decorate},
        decoration={markings,mark=at position .55 with {\arrow[draw=black]{>}}}},
	bigvector/.style={decorate, decoration={snake,amplitude=4pt}, draw},
}\usetikzlibrary{decorations.markings}

\tikzstyle{block} = [draw, rectangle, 
    minimum height=3em, minimum width=6em]

\newcommand{\Jonatan}[1]{{\color{red}#1}}
\newcommand{\Daniel}[1]{{\color{magenta}#1}}

\newcommand{\magenta}[1]{{\color{magenta}#1}}

\begin{document}

\title{\magenta{Phenomenology of an Extended $1+2$ Higgs Doublet Model with $S_3$ Family Symmetry}}
\author{A. E. C\'arcamo Hern\'andez}
\email{antonio.carcamo@usm.cl} 
\affiliation{Departamento de F\'isica, Universidad T\'{e}cnica Federico Santa Mar\'{\i}a, Casilla 110-V, Valpara\'{\i}so, Chile}
\affiliation{Centro Cient\'{\i}fico-Tecnol\'{o}gico de Valpara\'{\i}so,  Casilla 110-V, Valpara\'{\i}so, Chile}
\affiliation{Millennium Institute for Subatomic physics at high energy frontier - SAPHIR, Fernandez Concha 700, Santiago, Chile}
\author{Daniel Salinas-Arizmendi}
\email{daniel.salinas@usm.cl}
\affiliation{Departamento de F\'isica, Universidad T\'{e}cnica Federico Santa Mar\'{\i}a, Casilla 110-V, Valpara\'{\i}so, Chile}
\affiliation{Centro Cient\'{\i}fico-Tecnol\'{o}gico de Valpara\'{\i}so,  Casilla 110-V, Valpara\'{\i}so, Chile}
\author{Jonatan Vignatti}
\email{jonatan.vignatti@sansano.usm.cl}
\affiliation{Departamento de F\'isica, Universidad T\'{e}cnica Federico Santa Mar\'{\i}a, Casilla 110-V, Valpara\'{\i}so, Chile}
\affiliation{Escuela de Ingenier\'ia, Universidad Central de Chile, Avenida Francisco de Aguirre 0405, 171-0164 La Serena, Coquimbo, Chile}
\author{Alfonso Zerwekh}
\email{alfonso.zerwekh@usm.cl}
\affiliation{Departamento de F\'isica, Universidad T\'{e}cnica Federico Santa Mar\'{\i}a, Casilla 110-V, Valpara\'{\i}so, Chile}
\affiliation{Centro Cient\'{\i}fico-Tecnol\'{o}gico de Valpara\'{\i}so,  Casilla 110-V, Valpara\'{\i}so, Chile}
\affiliation{Millennium Institute for Subatomic physics at high energy frontier - SAPHIR, Fernandez Concha 700, Santiago, Chile}
\date{\today }

\begin{abstract}
In order to explain the mass hierarchy and mixing pattern in the leptonic sector, we explore an extension of the Standard Model whose scalar sector includes one active and two inert doublets as well as some scalar singlets. The model includes a $S_3$ family symmetry supplemented by extra cyclic symmetries. As a consequence of our construction, a Dark Matter (DM) candidate is predicted and its properties are consistent with the observed cosmic abundances and the constraints imposed by direct and indirect detection experiments.
The model allows Charged Lepton Flavor Violation (CLFV) processes like $\mu \rightarrow e\gamma$ and $\mu \rightarrow 3e$, but the predicted branching ratios align with experimental limits. 
Additionally, our analysis elucidates the generation of the active neutrino masses through a one-loop radiative seesaw mechanism matching the observed neutrino oscillation data. 
The model agrees with experimental data on Higgs Diphoton decay rates and on oblique parameters.

\end{abstract}
\maketitle

\section{Introduction}

The Standard Model (SM) of particle physics, which includes only one a Higgs Doublet for generating the masses of the elementary fermions and gauge bosons, provides a successfully minimal framework for understanding the fundamental interactions of subatomic particles. However, despite its successes, the SM leaves several fundamental questions unanswered, such as the nature of dark matter, the origin of neutrino masses and the pattern of lepton mixing. These limitations have motivated the development of 
extensions of the SM with discrete flavor symmetries whose spontaneous breaking produces the observed pattern of lepton masses and mixings. Among the different discrete flavor symmetries, the $S_3$ permutation symmetry is special popular, since it corresponds to the smallest non-abelian discrete group with a 
non-trivial singlet, and a doublet as irreducible representations. The $S_3$ flavor group corresponds to the symmetry group of equilateral triangle symmetry and has been implemented in extensions of the SM \cite{Gerard:1982mm,Kubo:2003iw,Kubo:2003pd,Kobayashi:2003fh,Chen:2004rr,Mondragon:2007af,Mondragon:2008gm,Bhattacharyya:2010hp,Dong:2011vb,Dias:2012bh,Meloni:2012ci,Canales:2012dr,Canales:2013cga,Ma:2013zca,Kajiyama:2013sza,Hernandez:2013hea,Ma:2014qra,Hernandez:2014vta,Hernandez:2014lpa,Vien:2014vka,Gupta:2014nba,CarcamoHernandez:2015mkh,Hernandez:2015zeh,Hernandez:2015hrt,Hernandez:2016rbi,CarcamoHernandez:2016pdu,Arbelaez:2016mhg,Gomez-Izquierdo:2017rxi,Cruz:2017add,Ma:2017trv,Espinoza:2018itz,Garces:2018nar,CarcamoHernandez:2018vdj,Gomez-Izquierdo:2018jrx,Pramanick:2019oxb,Garcia-Aguilar:2019chy,CarcamoHernandez:2020pxw,Garcia-Aguilar:2020vsy,Vien:2020trr,Espinoza:2020qyf,Gomez-Bock:2021uyu,Garcia-Aguilar:2021xgk,Gomez-Izquierdo:2023mph} to
produce a predictive pattern of lepton mixings consistent with the experimental data. The cobimaximal pattern for leptonic mixing is a good explanation for the measured neutrino oscillation experimental data. This pattern corresponds to a neutrino mass matrix of the form:
\begin{equation}
\widetilde{M}_{\nu}=\left( 
\begin{array}{ccc}
A & C & C^{*} \\ 
C & B & D \\ 
C^{*} & D & B^{*}
\end{array}%
\right),
\label{X}
\end{equation}
in the basis where the SM charged lepton mass matrix is diagonal. This pattern predicts $\theta_{13}\neq 0$, $\theta_{23}=\frac{\pi}{4}$ and $\delta_{CP}=-\frac{\pi}{2}$, which is well consistent with the experimental data on neutrino oscillations. The pattern is called cobimaximal because it predicts the maximal allowed leptonic mixing in the 23 plane as well as maximal leptonic Dirac CP violating phase. It also corresponds to a generalized $\mu-\tau$ symmetry ~\cite{babu:2002dz,grimus:2003yn,King:2014nza}: 
\begin{equation}
P^T\widetilde{M}_{\nu}P=\left(\widetilde{M}_{\nu}\right)^{*} 
\end{equation} 
with 
\begin{equation}
P=\left( 
\begin{array}{ccc}
1 & 0 & 0 \\ 
0 & 0 & 1 \\ 
0 & 1 & 0
\end{array}%
\right).
\label{X}
\end{equation}
To obtain the cobimaximal leptonic mixing pattern, non abelian discrete groups having triplets as irreducible representations such as, for instance $A_4$ \cite{Ma:2017moj,Ma:2021kfa} and $\Delta(27)$ \cite{Ma:2019iwj,CarcamoHernandez:2017owh,CarcamoHernandez:2018hst} have been used in extensions of the SM. In this work, we employ instead the $S_3$ flavor symmetry to generate a nearly cobimaximal leptonic mixing pattern, then making the model scalar sector more economical than the ones of cobimaximal mixing models based on the $A_4$ and $\Delta(27)$ family symmetries. Here we propose a $S_3$ flavored model with moderate amount of particle content, where the lepton mixing features a nearly cobimaximal mixing pattern, which is consistent with the observed neutrino oscillation experimental data. In the considered model, the scalar sector is composed of one active and two inert Higgs doublets ($1+2$ HDM), and several scalar singlets. The SM fermion sector is augmented by the inclusion of one right handed Majorana neutrino, thus allowing the implementation of a radiative seesaw mechanism for the generation of light active neutrino masses. In that theory, the SM gauge symmetry is supplemented by the inclusion of the non abelian discrete $S_3$ symmetry as well as some auxiliary cyclic symmetries. One of the auxiliary cylic symmetries, the $Z_2$ symmetry, is preserved, whereas the other discrete symmetries are spontaneously broken. The preserved $Z_2$ symmetry allows for a stable dark matter candidate and guarantees the radiative nature of the one loop level seesaw mechanism that produces the tiny active neutrino masses. The $S_3$ discrete symmetry as well as the other cyclic symmetries allow for a natural explanation of the SM charged lepton mass hierarchy as well as for a predictive and viable pattern of lepton mixings consistent with the neutrino oscillation experimental data. In this work, the spontaneous breaking of the $S_3$ discrete symmetry yield a controlled deviation of the cobimaximal mixing pattern of lepton mixings. 
In our case, there are two scalar doublets participating in the neutrino Yukawa interactions involved in the mass generation of light neutrinos that induce a one-loop radiative seesaw mechanism to generates an the tiny active neutrino masses. Such radiative seesaw mechanism is also mediated by a right handed Majorana neutrinos (see Refs.~\cite{Tao:1996vb,Ma:2006km,Cai:2017jrq,Arbelaez:2022ejo} for previous studies of this type of models). 


The content of this paper is as follows. Section \ref{model} presents the model and its details, such as symmetries, particle content, field assignments under the symmetry group, and describes the spontaneous symmetry breaking pattern. Section \ref{leptonsspectrum} discusses and analyzes the implications of the model on the masses and mixings in the lepton sector. Section \ref{scalar} describes the invariant scalar potential, the resulting scalar mass spectrum, and the mixing in the scalar sector. Section \ref{CLFV} provides an analysis of the charged lepton flavor violations. Additionally, the decay rate of the Higgs to two photons is studied in Section \ref{diphoton}. Section \ref{oblique} discusses the contribution of the model to the oblique parameters through the masses of the new scalar fields. In Section \ref{DM}, we study the scalar dark matter phenomenology. Finally, we state our conclusions in Section \ref{conclusion}.

\section{The model}\label{model}
Our proposed model corresponds to an extended $1+2$ Higgs Doublet Model, where the scalar sector is augmented by the inclusion of few singlet scalar fields, namely $\sigma$, $\chi$, $\rho$ and $\xi$, whereas the fermion sector is extended by adding one right handed Majorana neutrino, i.e., $N_R$. Furthermore, in the theory under consideration, the SM gauge symmetry is supplemented by the inclusion of the $S_{3}\otimes Z_{2}\otimes Z_{2}^{\prime} \otimes Z_{18}$ global symmetry, where the $S_{3}\otimes Z_{2}^{\prime} \otimes Z_{18}$ discrete group is spontaneously broken, whereas the $Z_{2}$ symmetry is preserved, thus preventing one-loop-level mass generation for active neutrinos, while allowing active neutrino masses to appear at one loop level. The successfull implementation of radiative seesaw mechanism that produces the tiny neutrino masses requires the inclusion of two dark $SU(2)$ scalar doublets, i.e., $H_1$, $H_2$ and one Majorana neutrino $N_R$, with non trivial charges under the preserved $Z_2$ symmetry. That radiative seesaw mechanism will be mediated by the neutral components of the $H_1$, $H_2$ scalar doublets as well as by the right handed Majorana neutrino $N_R$. The scalar and lepton content of the model with their transformations under the $S_{3}\otimes Z_{2}\otimes Z_{2}^{\prime} \otimes Z_{18}$ group are shown in Table \ref{leptons}, respectively, where the bold numbers correspond to the doublet, trivial, and non-trivial singlet representations for fields under $S_3$ symmetry. Given that the $SU(2)$ scalar doublets, i.e., $H_1$, $H_2$ have non trivial charges under the preserved $Z_2$ symmetry, they do not acquire vacuum expectation values, then forbidding masses for active neutrinos at tree level and allow them to appear at one loop level. Moreover, such preserved $Z_2$ symmetry guarantees the stability of the dark matter candidate, which will corresponds to the lightest of the particles having non trivial $Z_2$ charges. Furthermore, in order to generate the charged lepton mass hierarchy, we introduce the spontaneously broken $Z_{16}$ symmetry, whose spontaneous breaking by the vacuum expectation value of the scalar $\sigma$ whose spontaneous breaking produces the hierarchical structure of the charged lepton mass matrix crucial to yield the observed SM charged lepton mass pattern. Besides that, we consider the $S_3$ discrete symmetry since $S_3$ is the smallest non abelian group having a doublet and two singlets as irreducible representations as shown in the appendix \ref{appen:s3}. In our model we group the second and third families of left handed leptonic doublets in a $S_3$ double representation whereas the remaining SM leptonic fields are assign as $S_3$ singlets. This $S_3$ discrete group together with the spontaneously broken $Z_{2}^{\prime}$  gives rise to a nearly diagonal SM charged lepton mass matrix and to a light active neutrino mass matrix featuring a cobimaximal mixing pattern, with a moderate amount of particle content. Such nearly cobimaximal pattern of lepton mixings requires the inclusion of the $S_3$ doublets scalars $\xi$ and $\chi$ in the scalar spectrum of the model. This setup allows to generate a viable and predictive pattern of lepton mixings corresponding to a perturbed cobimaximal mixing pattern that allow to successfully accommodate the experimental values of the leptonic mixing angles and leptonic Dirac CP phase.  In our proposed model the full symmetry $\mathcal{G}$ exhibits the following spontaneous symmetry breaking pattern: 
\begin{eqnarray}
&&\mathcal{G}=SU\left( 3\right) _{C}\otimes SU\left( 2\right) _{L}\otimes
U\left( 1\right) _{Y}\otimes S_{3}\otimes Z_{2}\otimes Z_{2}' \otimes Z_{18}  \label{Group} \\
&&\hspace{35mm}\Downarrow \Lambda _{int}  \notag \\[3mm]
&&\hspace{15mm}SU\left( 3\right) _{C}\otimes SU\left( 2\right) _{L}\otimes
U\left( 1\right) _{Y}\otimes Z_{2}  \notag \\[3mm]
&&\hspace{35mm}\Downarrow \Lambda _{\text{EW}}  \notag \\[3mm]
&&\hspace{23mm}SU\left( 3\right) _{C}\otimes U\left( 1\right) _{em}\otimes
Z_{2}  \notag
\end{eqnarray}%
where the symmetry breaking scales satisfy the following hierarchy $\Lambda
_{int}\gg \Lambda _{\text{EW}}$, where $\Lambda _{\text{EW}}=246$ GeV. 
\begin{table}[]
\centering
\begin{tabular}{c c c c c c c c c c c c c c}
\toprule[0.2mm]
\hline
\vspace{0.1cm}   & $\ell_{1 L}$ &$ \ell_L$ & $\ell_{1 R}$ & $\ell_{2 R} $& $\ell_{3 R}$ &$ N_{R}$ & $\phi$ & $\chi$ & $\rho$ & $\sigma$ & $\xi$ & $H_1$ & $H_2$ \\  \hline\hline
 $S_3$ & $\mathbf{1}$ & $\mathbf{2}$ & $\mathbf{1}$ & $\mathbf{1}$ & $\mathbf{1}^{\prime}$ & $\mathbf{1}$ & $\mathbf{1}$ & $\mathbf{2}$ & $\mathbf{2}$ &$\mathbf{1}$ & $\mathbf{2}$ & $\bf{1}$ & $\bf{1}$\\
 $Z_2$ & $0$ & $0$ & $0$ & $0$ & $0$ & $1$ & $0$ & $0$ & $0$ & $0$ & $0$ & $1$ & $1$  \\
 $Z_2^{\prime}$ & $0$ & $0$ & $0$ & $1$ & $1$ & $0$ & $0$ & $0$ & $0$ &$0$ & $1$ & $0$ & $0$\\
 $Z_{18}$ & $-8$ & $-9$ & $1$ & $5$ & $-7$ & $0$ & $0$ & $0$ & $0$ & $-1$& $0$ & $9$ & $9$ \\\hline
\bottomrule[0.2mm]
\end{tabular}
\caption{Scalars and leptons assignments under the $S_3 \otimes Z_2 \otimes Z_2' \otimes Z_{18}$ discrete group.}
\label{leptons}
\end{table}

With the above particle content and symmetries specified in Table \ref{leptons}, 
the following relevant charged lepton Yukawa terms arise: 
\begin{eqnarray}
 \mathcal{L}_Y^{(\ell)} &=&Y_{11}^{(\ell)} \bar{\ell}_{1 L} \phi \ell_{1 R} \frac{\sigma^9}{\Lambda^9}+Y_{13}^{(\ell)} \bar{\ell}_{1 L} \phi  \ell_{3 R} \frac{\sigma\left(\xi\xi\xi\right)_\mathbf{1^{\prime}}}{\Lambda^4}+Y_{33}^{(\ell)} \left(\bar{\ell}_L \xi \right)_\mathbf{1'} \phi \ell_{3 R} \frac{ \sigma^2}{\Lambda^3}+Y_{22}^{(\ell)} \left(\bar{\ell}_L \xi \right)_\mathbf{1} \phi \ell_{2 R} \frac{\left(\sigma^*\right)^4}{\Lambda^5} \\ \nonumber
 & & 
 +\frac{1}{2}  m_{N} \overline{N}_{R} N_{R}^C 
\end{eqnarray}
whereas the neutrino Yukawa interactions take the form:
\begin{equation}\label{eq:leptonlagrangian}
   - \mathcal{L}_{Y}^{( \ell N)} =  \sum_{k=1}^2\left(y_{ k} \bar{\ell}_{L}  \widetilde{H}_k N_R \frac{\chi}{\Lambda}+
z_{ k} \bar{\ell}_{L}  \widetilde{H}_k N_R \frac{\rho}{\Lambda} 
   + x_{ k} \bar{\ell}_{1 L} \widetilde{H}_k N_R \frac{\sigma^*}{\Lambda}\right).
\end{equation}

Since the spontaneous breaking of the $Z_{16}$ symmetry produces the SM charged lepton mass hierarchy, we set the vacuum expectation values (VEVs),
as follows: 
\begin{equation}\label{vevs}
v_{\xi } \sim v_{\chi} \sim v_{\rho}=v_{\sigma } \sim \Lambda _{int} \sim \lambda \Lambda .
\end{equation}
where $\lambda=0.225$ is the Wolfenstein parameter and $\Lambda$ is the model cutoff which can be associated with the masses of the Froggatt-Nielsen messenger fields.

In order to generate a nearly cobimaximal pattern of lepton mixings, we consider the following VEV pattern for the $S_{3}$ doublet SM singlet
scalars $\xi$, $\chi$ and $\rho$ :
\begin{equation}
\left\langle \xi \right\rangle =v_{\xi }\left( 1, 0\right) ,\quad
\langle\chi\rangle=\frac{v_\chi}{\sqrt{2}}\left(e^{i \theta}, e^{-i \theta}\right), \quad \left\langle \rho \right\rangle = \frac{v_\rho}{\sqrt{2}}\left(1, 1\right)
\label{VEV}
\end{equation}
which is a natural solution of the scalar potential minimization equations for the whole region of parameter space, as shown in detail in Appendix \ref{decoupling}.

\section{Lepton masses and mixings}\label{leptonsspectrum}

\begin{figure}[]
\includegraphics[scale=0.4]{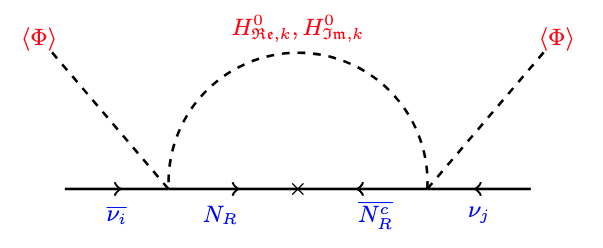}
\caption{Radiative see-saw that generates neutrinos masses via scalar component of the inert double, where $\Phi = \chi,\rho,\sigma$.}
\label{fig:radiativeloop}
\end{figure}

From the charged lepton Yukawa interactions in the Eq.~\eqref{eq:leptonlagrangian}, we find that the following SM charged lepton mass matrix arising after the SM gauge symmetry and the $S_3 \otimes Z_2' \otimes Z_{18}$ discrete group are spontaneously broken: 
\begin{equation}
  M_{\ell}= \frac{v}{\sqrt{2}}\left(\begin{array}{ccc}
a_1 \lambda^9 & 0 & b_1 \lambda^4 \\
0
& c_1 \lambda^5 & 0 \\
0 & 0 & a_2 \lambda^3
\end{array}\right)  
\end{equation}
where $a_1$, $a_2$, $b_1$ and $c_1$ are dimensionless parameters.  Concerning the neutrino sector, the preserved $Z_2$ symmetry prevents tree level masses for active neutrinos and allow the implementation of a radiative seesaw mechanism at one loop level that yields the active neutrino masses. Such one loop level radiative seesaw mechanism is mediated by the CP even and CP odd neutral components of the $SU(2)$ dark scalar doublets $H_1$ and $H_2$ as well as by the right handed Majorana neutrino $N_R$, as shown in the Feynman diagram of Figure \ref{fig:radiativeloop}.  Furthermore, the neutrino Yukawa terms yields the following mass matrix for light active neutrinos:
\begin{equation}
\widetilde{M}_{\nu} = \left(
\begin{array}{ccc}
A & C & C^{*} \\ 
C & B & D \\ 
C^{*} & D & B^{*}
\end{array}
\right)
\end{equation}

with $A$, $B$, and $C$ dimensionful parameters generated at one loop level which are given by the following relations:
\begin{eqnarray}
 A & \simeq &  \sum_{n=1}^2 m_N \lambda^2 x_{ n}^2 \mathcal{F}\left(m_{H^0_{\frak{Re},n}},m_{H^0_{\frak{Im},n}},m_N\right), \\   
 B  &\simeq& \sum_{n=1}^2 \frac{m_N \lambda^2}{2} \left[ \left(1+2\psi\right) y_{n}z_{n} + \psi y_{n}^2 z_{n}^2\right]  \mathcal{F}\left(m_{H^0_{\frak{Re},n}},m_{H^0_{\frak{Im},n}},m_N\right)e^{2 i\overline{\theta} },  \\ 
 C &\simeq& \sum_{n=1}^2  \frac{m_N \lambda^2}{\sqrt{2}} x_{n} \left( y_{n}+\psi z_{n}\right)  \mathcal{F}\left(m_{H^0_{\frak{Re},n}},m_{H^0_{\frak{Im},n}},m_N\right)e^{i \overline{\theta} },\\
 D & \simeq & \sum_{n=1}^2 \frac{m_N \lambda^2}{2} \left( \psi y_{n}+z_{n}\right)^2 \mathcal{F}\left(m_{H^0_{\frak{Re},n}},m_{H^0_{\frak{Im},n}},m_N\right)
\end{eqnarray}
where the loop function:
\begin{equation}
\mathcal{F}\left( m_1,m_2,m_3 \right) = \frac{1}{16\pi^2} \bigg[ \frac{m_1^2}{m_1^2-m_3^2} \ln\left(\frac{m_1^2}{m_3^2}\right) - \frac{m_2^2}{m_2^2-m_3^2} \ln\left(\frac{m_2^2}{m_3^2}\right)\bigg].
\end{equation}

\begin{table}[]
\begin{center} 
\begin{tabular}{c||c|c|c|c}
\toprule[0.2mm]
\hline
\multirow{2}{*}{\textbf{Observable}} & \multirow{2}{*}{\textbf{Model value}} & 
\multicolumn{3}{|c}{\textbf{Experimental value}} \\ \cline{3-5}
&  & $1\sigma $ range & $2\sigma $ range & $3\sigma $ range \\ \hline\hline
$m_{e}$ [MeV] & $0.489$ & $0.487$ & $0.487$ & $0.487$ \\ \hline
$m_{\mu }$ [MeV] & $102.9$ & $102.8\pm 0.0003$ & $102.8\pm 0.0006$ & $ 102.8\pm 0.0009$ \\ \hline
$m_{\tau }$ [GeV] & $1.75$ & $1.75\pm 0.0003$ & $1.75\pm 0.0006$ & $1.75\pm 0.0009$ \\ \hline
$\Delta m_{21}^{2}$ [$10^{-5}$eV$^{2}$]  & $7.47$ & $7.50_{-0.20}^{+0.22} $ & $7.11-7.93$ & $6.94-8.14$ \\ \hline
$\Delta m_{13}^{2}$ [$10^{-3}$eV$^{2}$]  & $2.55$ & $2.55_{-0.03}^{+0.02} $ & $2.49-2.60$ & $2.47-2.63$ \\ \hline
$\delta $ [$^{\circ }$]  & $209$ & $281_{-27}^{+23}$ & $229-328$ & $202-349$ \\ \hline
$\sin ^{2}\theta _{12}/10^{-1}$  & $ 3.13 $ & $3.18 \pm 0.16$ & $2.86-3.52$ & $2.71-3.69$ \\ \hline
$\sin ^{2}\theta _{23}/10^{-1}$  & $5.14$ & $5.74 \pm 0.14$ & $5.41-5.99$ & $4.34-6.10$ \\ \hline
$\sin ^{2}\theta _{13}/10^{-2}$  & $2.770$ & $2.200_{-0.062}^{+0.069}$ & 
$2.069-2.337$ & $2.000-2.405$ \\ \hline
\bottomrule[0.2mm]
\end{tabular}%
\end{center}
\caption{The model and experimental values of SM charged lepton masses, neutrino mass squared differences, leptonic mixing parameters and leptonic Dirac CP violating phase  
for the normal neutrino mass hierarchy (NH). The obtained values of the light active neutrino masses are also shown. The measured values for the charged lepton masses are taken from Ref. \cite{Xing:2020ijf}, whereas we use the experimental values of the neutrino mass squared differences, leptonic mixing parameterms and leptonic Dirac CP phase given in \cite{deSalas:2020pgw}.}
\label{Tab}
\end{table}

The experimental values of the SM charged lepton masses, the neutrino mass squared splittings and the leptonic mixing parameters can be successfully reproduced in terms of natural parameters of order one, as shown in Table \ref{Tab}, starting from the following benchmark point
\begin{equation}\label{eq:BS}
\begin{aligned}
a_1 \simeq & 1.957, \quad a_2 \simeq 0.859 \quad
b_1 \simeq 0.844 + 0.380i \quad
c_1 \simeq 1.0257, \\
A \simeq & -4.096 \ \text{meV}, \quad
B \simeq 39.04 \ \text{meV}, \quad
C \simeq -37.64 \ \text{meV}, \quad  D \simeq-34.25 \ \text{meV} , \quad \overline{\theta} \simeq -182.9^\circ .
\end{aligned}
\end{equation}
These values are obtained by solving the eigenvalue problem for the SM charged lepton and light active neutrino mass matrices. 
As shown in Table \ref{Tab} our model successfully describes the current SM charged lepton mass hierarchy as well as the neutrino oscillation experimental data. Using the obtained values of the $A$, $B$, and $C$ parameters, we can determine the right-handed Majorana neutrino mass $m_{N_R}$, based on the magnitudes of the light-active neutrino mass matrix elements From our numerical analysis we find that the heavy Majorana neutrino has a mass is of the order $1 \, \mathrm{TeV}$ and is heavier than the physical dark CP even and CP odd scalars arising from the neutral components of the $H_1$ and $H_2$ scalar doublets, whose masses acquire values at the subTeV scale. 

\begin{figure}
\centering
\includegraphics[scale=.3]{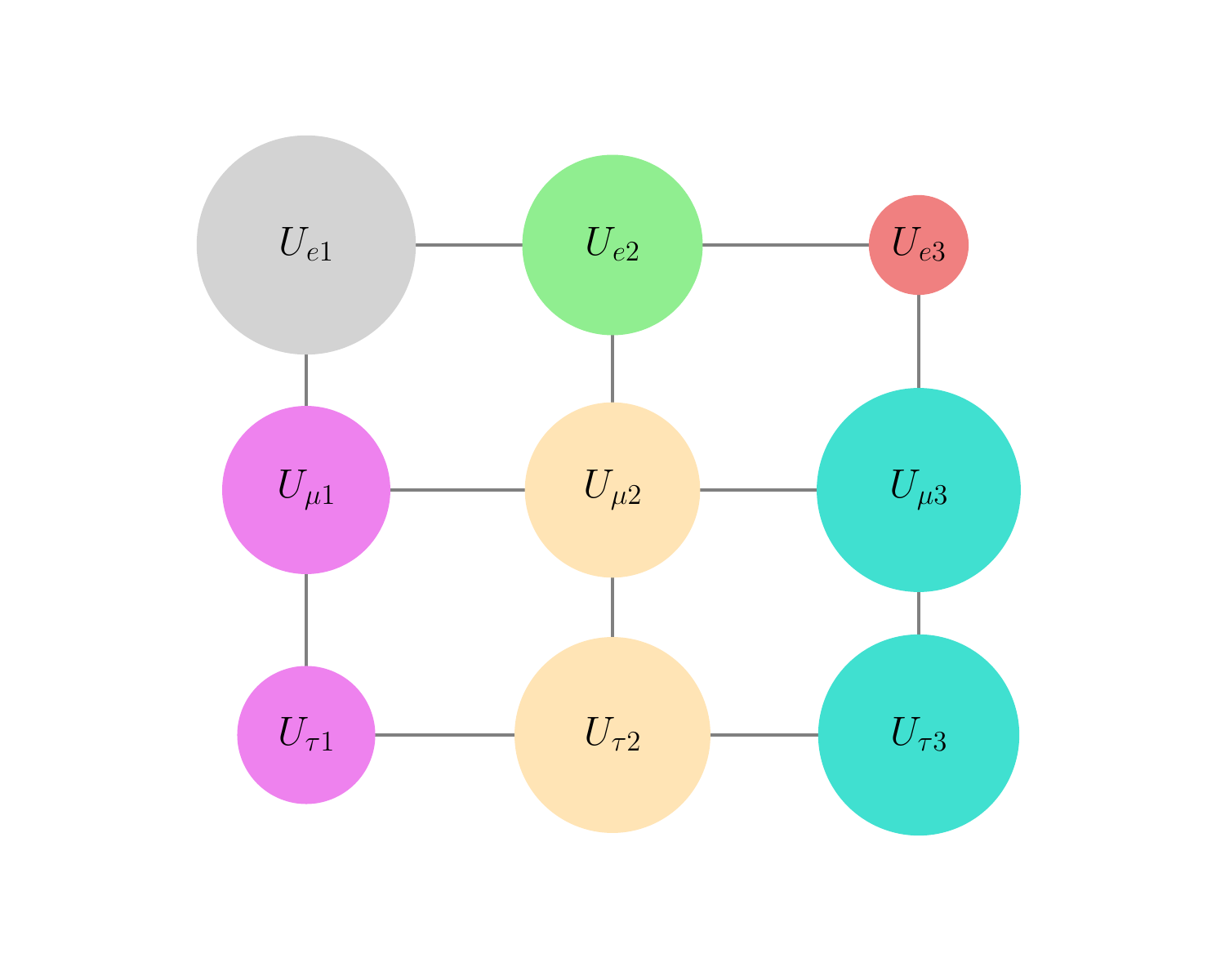}
\caption{Pictorial image illustrating the relative moduli of the PMNS matrix elements, where the size of the circle represents the proportion of the fit of our PMNS matrix values for our model.}
\label{fig:PMNS}
\end{figure}

A pictorial representation of the hierarchy in the lepton mixing angles of the Pontecorvo-Maki-Nakagawa-Sakata (PMNS) matrix fitted by our model, where the size of the modulus of the circle represents the modulus of the matrix entries, is shown in Figure~\ref{fig:PMNS}. Numerically using the values of the lepton model effective parameters of Eq.~\eqref{eq:BS}, the PMNS leptonic mixing matrix takes the form:
\begin{equation}
U_{\text{PMNS}} =R_{lL}^{\dagger} R_\nu=\left(
\begin{array}{ccc}
 -0.815-0.0515 i & -0.549+0.056 i & 0.102\, +0.132 i \\
 0.426\, -0.219 i & -0.463+0.238 i & 0.707\, -0.026 i \\
 -0.241-0.213 i & 0.609\, +0.231 i & 0.420\, +0.544 i \\
\end{array}
\right),
\end{equation}
where
\begin{eqnarray}
R_{lL} & = &  \left(
\begin{array}{ccc}
 0.972 & 0 & -0.235 \\
 0 & 0.912\, +0.410 i & 0 \\
 -0.215+0.097 i & 0 & -0.886+0.399 i \\
\end{array}
\right) ,  \\
R_\nu & = &  \left(
\begin{array}{ccc}
 -0.736 & -0.677 & 0 \\
 0.478\, -0.025 i & -0.520+0.027 i & 0.655\, +0.267 i \\
 0.478\, +0.025 i & -0.520-0.027 i & -0.624-0.333 i \\
\end{array}
\right).
\end{eqnarray}

\section{The Scalar Sector}\label{scalar}
\label{scalarsector}
\begin{figure}[]
\centering
\includegraphics[scale=.65]{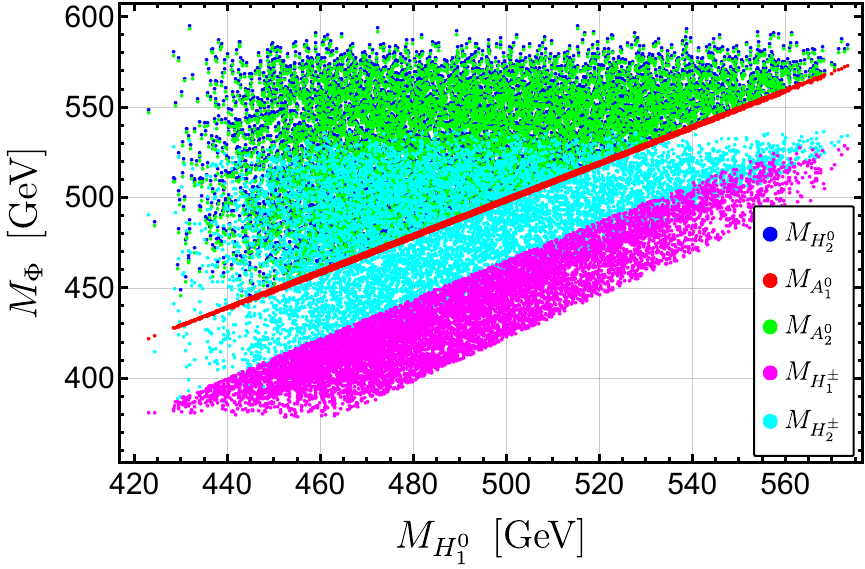}
\caption{The physical scalar mass correlation, $M_\Phi$ corresponde a $M_{H_2^0}$ (blue point),$M_{A_1^0}$ (red point), $M_{A_2^0}$ (green point), $M_{H_1^\pm}$ (magenta point) and $M_{H_2^\pm}$ (cyan point).}
\label{mass:correlationscalar}
\end{figure}

In this section we proceed to analyze the low energy scalar potential of the model. Given that we are considering that the discrete $S_{3}\otimes Z_{2}' \otimes Z_{18}$ group is spontaneously broken at a scale much larger than the electroweak symmetry breaking scale, the gauge singlet scalars of the model acquire very large vacuum expectation values (VEVs), then yielding very small mixing angles of these fields with the CP even part of the SM Higgs doublet. It is worth mentioning that these mixing angles are very small since they are suppressed by the ratios of their VEVs (assumed that the quartic scalar couplings are of the same order of magnitude), which is a consequence of the method of recursive expansion proposed in \cite{Grimus:2000vj}. Consequently, the couplings of the 126 GeV SM-like Higgs boson with SM particles are very close to the SM expectation, then implying that the alignment is naturally fulfilled in our model. Therefore the singlet scalar fields can be decoupled in the low energy phenomenology, where the only relevant scalar interactions are the ones arising from the 
potential of the two inert Higgs doublets and the SM Higgs doublet, which is given by
\begin{eqnarray} \label{potentialHiggs}
\nonumber V(\phi, H_{1}, H_{2}) &=& m_1^2 \phi^{\dagger} \phi+m_2^2 H_{1}^{\dagger} H_{1}+ m_3^2 H_{2}^{\dagger} H_{2} + m_4^2 \left( H_{1}^{\dagger} H_{2} + H.c. \right) +\frac{1}{2} \lambda_1\left(\phi^{\dagger} \phi\right)^2+\frac{1}{2} \lambda_2\left(H_{1}^{\dagger} H_{1}\right)^2 \\
& & +\frac{1}{2} \lambda_3\left(H_{2}^{\dagger} H_{2}\right)^2 + \lambda_4\left(\phi^{\dagger} \phi\right)\left(H_{1}^{\dagger} H_{1}\right)+\lambda_5\left(\phi^{\dagger} \phi\right)\left(H_{2}^{\dagger} H_{2}\right)+\lambda_6\left(H_{1}^{\dagger} H_{1}\right)\left(H_{2}^{\dagger} H_{2}\right)  \nonumber \\
& & + \lambda_7\left(\phi^{\dagger} H_{1}\right)\left(H_{1}^{\dagger} \phi\right)+\lambda_8\left(\phi^{\dagger} H_{2}\right)\left(H_{2}^{\dagger} \phi\right)+\lambda_9\left(H_{1}^{\dagger} H_{2}\right)\left(H_{2}^{\dagger} H_{1}\right)   \\
& & + \frac{1}{2} \lambda_{10}\left[\left(\phi^{\dagger} H_{1}\right)^2+\right. H.c. ]
+\frac{1}{2} \lambda_{11}\left[\left(\phi^{\dagger} H_{2}\right)^2+\right. H.c. ]
+\frac{1}{2} \lambda_{12}\left[\left(H_{1}^{\dagger} H_{2}\right)^2+ H.c. \right] , \nonumber
\end{eqnarray}
where the quartic scalar couplings are taken to be real since we assume a CP conserving scalar potential. Furthermore, $\phi$ is the Standard Model doublet whereas $H_1$ and $H_2$ are dark scalar doublets. These $H_1$ and $H_2$ scalar doublets mix and their mixing is controlled by the $m_4$ parameter. Notice that thanks to the preserved $Z_2$ symmetry, under which the dark $H_k (k=1,2)$ scalar doublets are charged, their components do not have tree level mixings with the ones of the SM Higgs doublet. 
On the other hand, it is worth mentioning that the parameters of the potential in Eq.(\ref{potentialHiggs}) are subject to various constraints. One of these constraints arises from the perturbative unitarity which yields the bound $|\lambda_i|< 4\pi$ for the $\lambda_i$ quartic scalar couplings. The other one corresponds to the requirement that the scalar potential must be bounded from below, in order to have a stable vacuum. To determine the bounded-from-below conditions we extend the approach of \cite{Bhattacharyya:2015nca} to our model. In such approach we analyze the behavior of the quartic terms since these terms will determine the shape of the scalar potential in the region of very large values of the field components. The quartic terms of the scalar potential can be written as follows:
\begin{eqnarray}
    V_4 &=& \frac{1}{4} \left[ \left(\sqrt{\lambda_1} a - \sqrt{\lambda_2} b \right)^{2} + \left(\sqrt{\lambda_1} a - \sqrt{\lambda_3} c \right)^{2} + \left(\sqrt{\lambda_2} b - \sqrt{\lambda_3} c \right)^{2} \right] \nonumber \\
    &&+ \left( \frac{1}{2} \sqrt{\lambda_1 \lambda_2} + \lambda_4  \right) \left( a b - d^{2}  \right) + \left( \frac{1}{2} \sqrt{\lambda_1 \lambda_3} + \lambda_5  \right) \left( a c - g^{2}  \right) + \left( \frac{1}{2} \sqrt{\lambda_2 \lambda_3} + \lambda_6  \right) \left( b
 c - k^{2}  \right)  \\
    && + \left( \lambda_4 + \lambda_7 + 2 \sqrt{\lambda_1 \lambda_2} \right) d^{2} + \left( \lambda_{10} - \sqrt{\lambda_1 \lambda_2} \right) \left( d^{2} + e^{2} \right) + \left( \lambda_7 - \lambda_{10} - \sqrt{\lambda_1 \lambda_2} \right) e^{2} \nonumber \\
    && + \left( \lambda_5 + \lambda_8 + 2 \sqrt{\lambda_1 \lambda_3} \right) g^{2} + \left( \lambda_{11} - \sqrt{\lambda_1 \lambda_3} \right) \left( g^{2} + f^{2} \right) + \left( \lambda_8 - \lambda_{11} - 
\sqrt{\lambda_1 \lambda_3} \right) f^{2} \nonumber \\
    && + \left( \lambda_6 + \lambda_9 + 2 \sqrt{\lambda_2 \lambda_3} \right) k^{2} + \left( \lambda_{12} - \sqrt{\lambda_2 \lambda_3} \right) \left( h^{2} + k^{2} \right) + \left( \lambda_9 - \lambda_{12} - \sqrt{\lambda_2 \lambda_3} \right) h^{2}, \nonumber 
\end{eqnarray}
with
\begin{align}
\nonumber a &\equiv \phi^{\dagger} \phi, \quad b \equiv H_1^{\dagger} H_1, \quad c \equiv H_2^{\dagger} H_2, \\
d &\equiv \mathfrak{Re} (\phi^{\dagger} H_1), \quad e \equiv \mathfrak{Im} (\phi^{\dagger} H_1), \quad f \equiv \mathfrak{Re} (\phi^{\dagger} H_2), \\
\nonumber g &\equiv \mathfrak{Im} (\phi^{\dagger} H_2), \quad h \equiv \mathfrak{Re} (H_1^{\dagger} H_2), \quad k \equiv \mathfrak{Im} (H_1^{\dagger} H_2).
\end{align}
Then following the procedure used for analyzing the stability described in Refs. \cite{Bhattacharyya:2015nca,Maniatis:2006fs}, we find that the following bounded-from-below conditions of the scalar potential: 
\begin{gather}
\lambda_1, \lambda_2,\lambda_3>0,\quad  \sqrt{\lambda_1 \lambda_2} > -2\lambda_4, \lambda_{10}, \lambda_7-\lambda_{10} ,-\frac{1}{2}\left(\lambda_4+\lambda_7\right), \notag\\
 \sqrt{\lambda_1 \lambda_3} > -2\lambda_5 ,\lambda_{11}, \lambda_8-\lambda_{11},-\frac{1}{2} \left(\lambda_5+\lambda_8 \right),\\
 \sqrt{\lambda_2 \lambda_3} > -2\lambda_6 ,\lambda_{12}, \lambda_9-\lambda_{12},-\frac{1}{2} \left(\lambda_6+\lambda_9 \right).\notag 
\end{gather}
From the analysis of the scalar potential we find after the spontaneous symmetry breaking, that the scalar spectrum is composed of a light CP even scalar state $h$ arising from $\phi$ corresponding to the $126$ SM like Higgs boson with mass $M_{h} = \sqrt{2\lambda _1}v$, three massless states which correspond to the Goldstone bosons $G_1^{\pm}$, $G_1^{0}$ also arising from $\phi$ and related to the longitudinal components of SM $W^{\pm}$ and $Z$ gauge bosons, respectively, as well as well two CP even, two CP odd and four electrically charged non-SM scalars coming from the dark $H_1$ and $H_2$ $SU(2)$ scalar doublets. We also find that the non SM scalar masses are given by:
\begin{eqnarray}
M_{H_{1,2}^{\pm}}^2 &=& \frac{m_2^2+m_3^2}{2} + \frac{1}{4} \left(v^2\mathfrak{Im}\lambda_a\mp \sqrt{\left[2 m_2^2-2 m_3^2+ v^2\mathfrak{Re}\lambda_a \right]^2+16 m_4^4}\right),\\
M_{A_{1,2}^{0}}^2 &=& \frac{m_2^2+m_3^2}{2}+ \frac{1}{4} \left(v^2\lambda _{c}\mp \sqrt{\left(2 m_2^2-2 m_3^2+ v^2\lambda_b\right){}^2+16 m_4^4}\right),\\
M_{H_{1,2}^{0}}^2 &=&\frac{m_2^2+m_3^2}{2}+  \frac{1}{4} \left(v^2\lambda_e \mp \sqrt{\left(2 m_2^2-2 m_3^2+v^2\lambda_d \right)^2+16 m_4^4} \right),
\end{eqnarray}
where we have defined
\begin{eqnarray}
\lambda_a &=& \left(\lambda_4-\lambda_5\right)+i \left(\lambda_4+\lambda_5\right),\\
\lambda_{b,c} &=&\lambda_4\mp \lambda _5+\lambda _7 \mp\lambda _8-\lambda_{10}\pm\lambda_{11} ,\\
\lambda_{d,e} &=&\lambda_4\mp \lambda _5+\lambda _7 \mp\lambda _8+\lambda _{10}\mp \lambda_{11}.
\end{eqnarray}
Here, the subscript $H_{1,2}^{\pm}$, $A_{1,2}^{0}$, and $H_{1,2}^{0}$ denote the charged scalar, pseudoscalar, and scalar neutral fields respectively. From our numerical analysis we have found 
that the dark heavy non SM scalars have a quasidegenerate mass spectrum. Figure \ref{mass:correlationscalar}, displays the correlations of the mass $H_1 ^{0}$ with the other non SM scalar masses, where the best fit point corresponds to the central values 

\begin{equation}
m_2\simeq 385.5\ \text{GeV},\ m_3\simeq 388.5\ \text{GeV} \ \text{and} \ m_4\simeq 2.6\ \text{GeV}.
\end{equation}

Furthermore, we find that the dark scalars in mass and interaction basis are related by: 
\begin{equation} \label{physical-fields}
\begin{array}{ccc}
H_1^\pm = C_\beta  h_1^\pm +S_\beta h_2^\pm,  \quad  & A_1^0 = C_\gamma  \zeta_1 +S_\gamma \zeta_2, \quad & H_1^0 = C_\alpha  \xi_1 +S_\alpha \xi_2,\\
H_2^\pm = -S_\beta  h_1^\pm +C_\beta h_2^\pm, & A_2^0 = -S_\gamma  \zeta_1 +C_\gamma \zeta_2, & H_2^0 = -S_\alpha  \xi_1 +C_\alpha \xi_2,
\end{array}
\end{equation}
where we have used a reduced notation for the trigonometric functions: $C_i=\cos(i)$, $S_i=\sin(i)$, and $i=\alpha,\beta,\gamma$. Here the $H_{1,2}^0$, $A_{1,2}^0$ and $H_{1,2}^{\pm}$ fields correspond to the CP-odd, CP-even and electrically charged scalars in the interaction basis, respectively. Besides that, the 
mixing angles are given by:
\begin{eqnarray}
\tan 2 \alpha &= & \frac{4 m_4^2}{\lambda_d v^2 +2m_2^2-2m_3^2},\\
\tan 2 \beta &=& \frac{4 m_4^2}{\text{Re}\lambda_a v^2 +2m_2^2-2m_3^2},\\
\tan 2 \gamma &=& \frac{4 m_4^2}{\lambda_e v^2 +2m_2^2-2m_3^2}.
\end{eqnarray}
The extensions of the scalar sector in the HDM models include new fields that introduce important phenomenological studies, e.g. in our $1+2$ model, the extra charged fields introduce loop corrections in the photon production from the decay of the SM-type Higgs signal, precisely measured by the CMS \cite{Saha:2022cnz} and ATLAS \cite{ATLAS:2022tnm} collaborations. Similarly, these charged fields, together with the even and odd CPs, can be used to determine the constraints introduced by the slant parameters $S$, $T$, and $U$, which come from the study of the propagator corrections of the model's vector particles; these parameters provide strong constraints on the BSM. In the following sections we discuss several cases where even and odd neutral and charged CP scalars have a direct impact, such as the charged lepton flavor violation problem present in the current Standard Model, the Higgs diphoton decay, the oblique parameter corrections, and finally a good scalar dark matter candidate constrained within the experimental data and future predictions as \cite{XENON:2018voc,DARWIN:2016hyl}.


\section{Charged lepton flavor violation (CLFV)}\label{CLFV}

Neutrino flavor oscillations suggest that leptonic flavor is not a conserved quantity in nature, as predicted by the Standard Model. Currently, charged lepton flavor violating (CLFV) processes have not been directly observed, but it is expected that at higher energies we may observe flavor violations in charged leptonic interactions, as occurs in the neutrino flavor.  The most stringent limits for CLFV arises from muon decay measurements, namely $\mu \rightarrow e \gamma$, in the latter experimental result sets an upper bound on the branching ratio, which reads as $\text{Br} \left( \mu \rightarrow e \gamma \right) < 4.2 \times 10^{-13}$ \cite{MEG:2016leq}, this upper bound on the branching ratio for 
$\mu \rightarrow e \gamma$ strongly constrain the parameter space of beyond standard model theories. In this section we will focus on the CLFV process involving charged leptons. The CLFV decay processes arise via one-loop diagrams mediated by masses $M_{H_{1,2}^{\pm}}$ of the two charged scalar components of the $SU(2)_L$ inert doublet the $H_{1,2}$ and $M_{N_R}$ correspond to the masses of  the right-handed Majorana neutrino $N_R$ particles. The Branching ratio for $\ell_\alpha \rightarrow \ell_\beta \gamma$ and the branching ratio for $3$-body decays $\ell_\alpha \rightarrow 3\ell_\beta$ respectively, where $\alpha,\beta = e, \mu, \tau$ is provided by \cite{Toma:2013zsa,Vicente:2014wga,Abada:2022dvm,Hernandez:2021iss}:

\begin{eqnarray}
\text{Br}\left(\ell_\alpha \rightarrow \ell_\beta \gamma\right)   & = & 
\frac{3(4 \pi)^3 \alpha_{\mathrm{EM}}}{4 G_F^2}\left|A_D\right|^2 \mathrm{Br}\left(l_\alpha \rightarrow l_\beta \nu_\alpha \overline{\nu_\beta}\right)
, \\
\text{Br}\left(\ell_\alpha \rightarrow \ell_\beta \overline{\ell_\beta} \ell_\beta\right) &= & \frac{3(4 \pi)^2 \alpha_{\mathrm{EM}}^2}{8 G_F^2}\left[\left|A_{N D}\right|^2+\left|A_D\right|^2\left(\frac{16}{3} \log \left(\frac{m_\alpha}{m_\beta}\right)-\frac{22}{3}\right)+\frac{1}{6}|B|^2\right. \\
& &  \left.+\left(-2 A_{N D} A_D^*+\frac{1}{3} A_{N D} B^*-\frac{2}{3} A_D B^*+H.c.\right)\right] \text{Br}\left(\ell_\alpha \rightarrow \ell_\beta \nu_\alpha \overline{\nu_\beta}\right),
\end{eqnarray}
where the form factors $A_D$ and $A_{ND}$ come from the dipole-photon penguin diagrams and the non-dipole photon penguin diagrams, and $B$ is the contribution from the box diagrams, they can be expressed as
\begin{eqnarray}
A_D & = &  \ \sum_{k=1}^2 \frac{\widetilde{x}_{\alpha,k} \widetilde{x}_{\beta,k}^{*}}{2(4 \pi)^2} \frac{1}{M_{H_{k}^{ \pm}}^2} F_2\left(
\frac{M_{N_R}^2}{M_{H_{k}^{ \pm}}^2}
\right),\\
A_{N D} & =& \  \sum_{k=1}^3 \frac{\widetilde{x}_{\alpha,k} \widetilde{x}_{\beta,k}^{*}}{6(4 \pi)^2} \frac{1}{M_{H_{k}^{ \pm}}^2} G_2\left(\frac{M_{N_R}^2}{M_{H_{k}^{ \pm}}^2}\right),\\
 \label{eq:boxdiagram} e^2 B & = & \ \frac{1}{(4 \pi)^2 M_{H_{k}^{ \pm}}^2} \sum_{k=1}^2 \left[\frac{1}{2}
\left\vert \widetilde{x}_{\beta,k}\right\vert^2 \widetilde{x}_{\beta,k}^* \widetilde{x}_{\alpha,k}
\overline{D}_1\left( \frac{M_{N_R}^2}{M_{H_{k}^{ \pm}}^2}\right) 
+\frac{M_{N_R}^2}{M_{H_{k}^{ \pm}}^2} \left\vert \widetilde{x}_{\beta,k} \right\vert^2 \widetilde{x}_{\beta,k}^* \widetilde{x}_{ \alpha,k}
\overline{D}_2\left(\frac{M_{N_R}^2}{M_{H_{k}^{ \pm}}^2}\right)\right].
\end{eqnarray}
where $\widetilde{x}_{\delta,k}= \sum_{i=1}^3 \widetilde{y}_{i,k}\left(R_{\ell L}^{\dagger}\right)_{\delta i}$, with $R_{\ell L}$ being the left-handed charged lepton mixing matrix, $G_F$ is the Fermi constant, $\alpha_{\text{em}}$ is the electromagnetic fine structure constant.

The different loop functions take the form
\begin{equation}
\begin{aligned}
F_2(x) & =\frac{1-6 x+3 x^2+2 x^3-6 x^2 \log x}{6(1-x)^4}, \\
G_2(x) & =\frac{2-9 x+18 x^2-11 x^3+6 x^3 \log x}{6(1-x)^4}, \\
D_1(x, y) & =-\frac{1}{(1-x)(1-y)}-\frac{x^2 \log x}{(1-x)^2(x-y)}-\frac{y^2 \log y}{(1-y)^2(y-x)}, \\
D_2(x, y) & =-\frac{1}{(1-x)(1-y)}-\frac{x \log x}{(1-x)^2(x-y)}-\frac{y \log y}{(1-y)^2(y-x)} .
\end{aligned}
\end{equation}
where  we have defined $\overline{D}_k(x) = \lim _{y\rightarrow x} D_k\left(x,y \right)$ in the Eq.~\eqref{eq:boxdiagram}, since loop functions are obtained for several neutrino mediators.

Another leptonic flavour violation process, is the conversion $\mu-e$ in the nuclei, whose ratio $\mu^- -e^-$ is defined as \cite{Lindner:2016bgg}
\begin{equation}
\text{CR}(\mu-e)=\frac{\Gamma\left(\mu^{-}+\operatorname{Nucleus}(A, Z) \rightarrow e^{-}+\operatorname{Nucleus}(A, Z)\right)}{\Gamma\left(\mu^{-}+\operatorname{Nucleus}(A, Z) \rightarrow \nu_\mu+\operatorname{Nucleus}(A, Z-1)\right)}.
\end{equation}

For the radiative neutrino mass model considered in this work, the conversion rate, normalized to the charged lepton capture rate, takes the form Ref.~\cite{Vicente:2014wga}:
\begin{equation}
\begin{aligned}
\text{CR}\left(\ell_\alpha N \rightarrow \ell_\beta N\right)= & \ \frac{p_\beta E_\beta m_\alpha^3 G_F^2 \alpha_{\mathrm{EM}}^3 Z_{\text {eff }}^4 F_p^2}{8 \pi^2 Z \Gamma_{\text {capt }}}\left[\left|(Z+N)\left(g_{L V}^{(0)}+g_{L S}^{(0)}\right)+(Z-N)\left(g_{L V}^{(1)}+g_{L S}^{(1)}\right)\right|^2\right. \\
& \left.+\left|(Z+N)\left(g_{R V}^{(0)}+g_{R S}^{(0)}\right)+(Z-N)\left(g_{R V}^{(1)}+g_{R S}^{(1)}\right)\right|^2\right],
\end{aligned}
\end{equation}
where the expressions for the quantities $Z_{\text {eff }}, F_p, \Gamma_{\text {capt }}$, and $g_{L / R S / V}^{(i)}$ are given in Refs.~\cite{Arganda:2007jw}.

\begin{figure}
\centering
\subfigure[]{\includegraphics[scale=0.6]{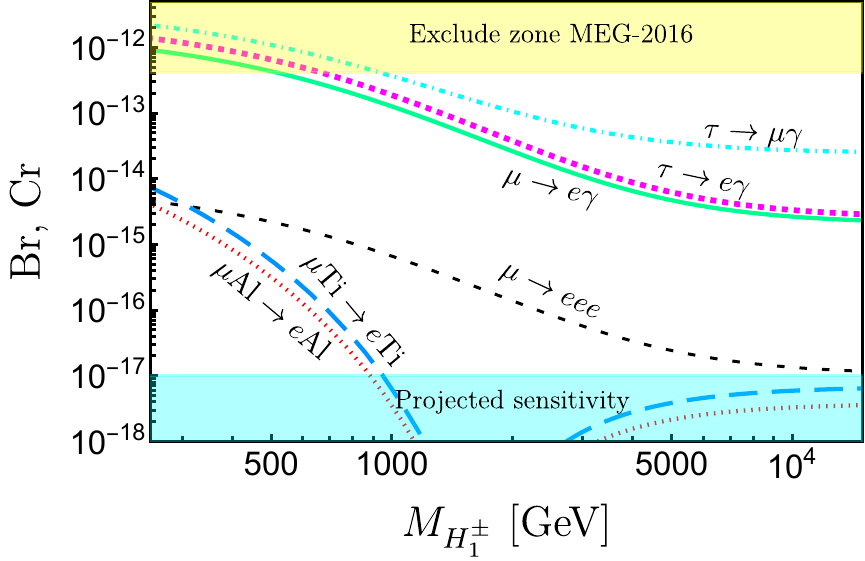}}
\subfigure[]{\includegraphics[scale=0.6]{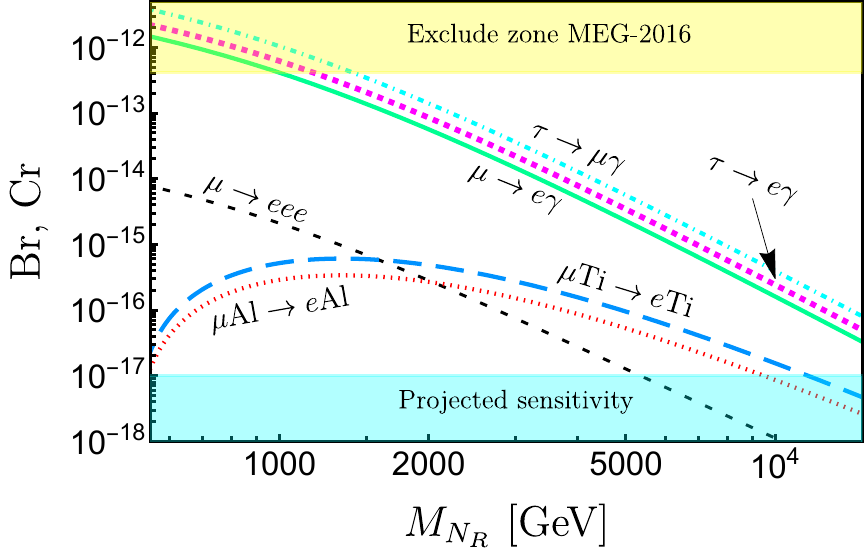}}
\caption{The Branching ratio for $\mu \rightarrow e \gamma$, $\tau \rightarrow \mu \gamma$, $\gamma \rightarrow e \gamma$ and $\mu\rightarrow eee$, and the $\mu-e$ conversion in the nuclei of Ti and Al as a function of the mass of the (a) charged scalar $H_1^{\pm}$ (with $M_{N_R}=992\ \text{GeV}$ and $M_{H_2^{\pm}}=365\ \text{GeV}$), (b) right-handed Majorana neutrino $N_R$ (with $M_{H_1^{\pm}}=511\ \text{GeV}$ and $M_{H_2^{\pm}}=365\ \text{GeV}$). The upper shaded region corresponds to an area excluded by the MEG 
\cite{MEG:2016leq} and the lower shaded region corresponds to the expected sensitivities of the next generation of experiments using aluminium as targets \cite{Bernstein:2013hba}.}
\label{fig:plotCLFV}
\end{figure}

In the Figure~\ref{fig:plotCLFV} shows the mass dependence of the charged scalar field $H_1^\pm$ and the right-handed Majorana neutrino $N_R$ on the branching ratios $\mu \rightarrow e \gamma$, $\tau \rightarrow \mu \gamma$ and $\gamma \rightarrow e \gamma$, and for the three-body decay $\mu \rightarrow 3e$, note that for our model the curves of $\mu \rightarrow e \gamma$ and $\mu \rightarrow 3e$ are correlated due to the smallness of the effective yukawas, in the upper region of the current bound \cite{MEG:2016leq}, which is the strongest constraint coming from the $\mu \rightarrow e \gamma$ process, as noted earlier. The same graph shows the curves for $\mu-e$ conversion in the nuclei of titanium and aluminium, where the current projected sensitivity $\text{CR}\left(\mu^{-} \text{Al} \rightarrow e^{-} \text{Al} \right) \lesssim 10^{-17}$ \cite{Bernstein:2013hba} is included by the lower shaded region.

\section{The Higgs Diphoton decay rate constraints}
\label{diphoton}
\begin{figure}[H]
\centering
\includegraphics[scale=.45]{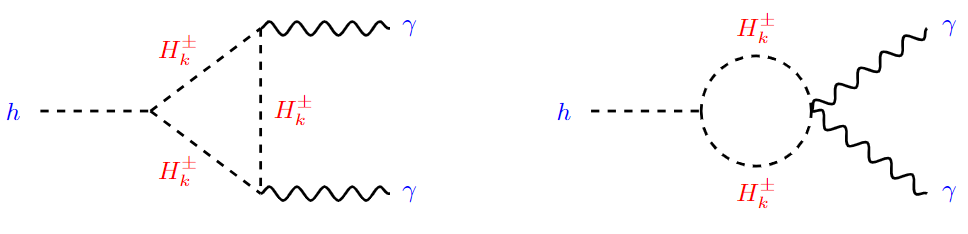}
\caption{Extra one-loop Feynman digrams in the unitary gauge contributing to the higgs diphoton decay.}
\label{feyndiagr1}
\end{figure}

In this section we analyze the implications of the model in the SM Higgs decay into a photon pair. In the standard model, the decay of the Higgs into two photons receives contributions arising from the $W$ boson and $t$ quark loops. In our model, we must add the extra contributions from loops with charged scalars $H_k^{\pm}$, as shown in Figure~\ref{feyndiagr1}. 
The explicit form the decay rate for the $h\rightarrow \gamma \gamma$ process takes the for \cite{Bhattacharyya:2014oka,Logan:2014jla,Hernandez:2021uxx}
\begin{equation}
\Gamma(h \rightarrow \gamma \gamma)=\frac{\alpha_{\text{em}}^2 m_h^3}{256 \pi^3 v^2}\left|\sum_f \mathcal{A}_{hff} N_C Q_f^2 F_{\frac{1}{2}}\left(\varrho_f\right)+\mathcal{A}_{hWW} F_1\left(\varrho_W\right)+\sum_{k=1}^2 \mathcal{A}_{h H_k^{\pm} H_k^{\mp} } F_0\left(\varrho_{H_k^{\pm}}\right)\right|^2
\end{equation}

where $\varrho_i$ are the mass ratios $\varrho_i=\frac{m_h^2}{4 M_i^2}$ with $M_i=m_f, M_W, M_{H_{1,2}
^{\pm}}$, $\alpha_{\text{em}}$ is the fine structure constant; $N_C$ is the color factor ($N_C=1$ for leptons and $N_C=3$ for quarks) and $Q_f$ is the electric charge of the fermion in the loop. From the fermion-loop contributions we only consider the dominant top quark term. Furthermore, $\mathcal{A}_{h H_k^{\pm} H_k^{\mp}}=\frac{\mathcal{C}_{h H_k^{\pm} H_k^{\mp}}v}{M_{H_k^{\mp}}^2}$ being $\mathcal{C}_{h H_k^{\pm} H_k^{\mp}}$ the trilinear adimensional coupling between the SM-like Higgs and a charged Higgs boson pair 
whereas $\mathcal{A}_{htt}$ and $\mathcal{A}_{hWW}$ are the deviation factors from the SM Higgs-top quark coupling and the SM Higgs-W gauge boson coupling, respectively (in the SM the factors $\mathcal{A}_{htt}$ is normalized). Such deviation factors are close to unity in our model, which is a consequence of the numerical analysis of its scalar, Yukawa and gauge sectors.

Furthermore, $F_{\frac{1}{2}}(\varrho)$ and $F_1(\varrho)$ are the dimensionless loop factors for spin-1/2 and spin-1 particles running in the internal lines of the loops. They are given by:
\begin{eqnarray}
F_0(\varrho	) & =& -\varrho(1-\varrho f(\varrho)),\\
F_{\frac{1}{2}}(\varrho) & =& 2\varrho(1+(1-\varrho) f(\varrho)), \\
F_1(\varrho) & =& -\left(2+3\varrho+3\varrho \left(2-\varrho\right) f(\varrho)\right),
\end{eqnarray}
with

\begin{equation}
f(\varrho)= \begin{cases}\arcsin ^2 \sqrt{\varrho^{-1}} & \text { for } \varrho \geq 1 \\ -\frac{1}{4}\left[\ln \left(\frac{1+\sqrt{1-\varrho}}{1-\sqrt{1-\varrho}}\right)-i\pi\right] ^2& \text { for } \varrho<1\end{cases}
\end{equation}

In order to study the implications of our model in the decay of the $126\ \mathrm{GeV}$ Higgs into a photon pair, one introduces the Higgs diphoton signal strength $R_{\gamma \gamma}$, which is defined as \cite{Bhattacharyya:2014oka}:
\begin{equation}
R_{\gamma \gamma}=\frac{\sigma(p p \rightarrow h) \Gamma(h \rightarrow \gamma \gamma)}{\sigma(p p \rightarrow h)_{\text{SM}} \Gamma(h \rightarrow \gamma \gamma)_{\text{SM}}} \simeq \mathcal{A}_{h t t}^2 \frac{\Gamma(h \rightarrow \gamma \gamma)}{\Gamma(h \rightarrow \gamma \gamma)_{\text{SM}}} .
\end{equation}

That Higgs diphoton signal strength, normalizes the $\gamma \gamma$ signal predicted by our model in relation to the one given by the SM. Here we have used the fact that in our model, single Higgs production is also dominated by gluon fusion as in the Standard Model.
\begin{figure}[]
\centering
\includegraphics[scale=.5]{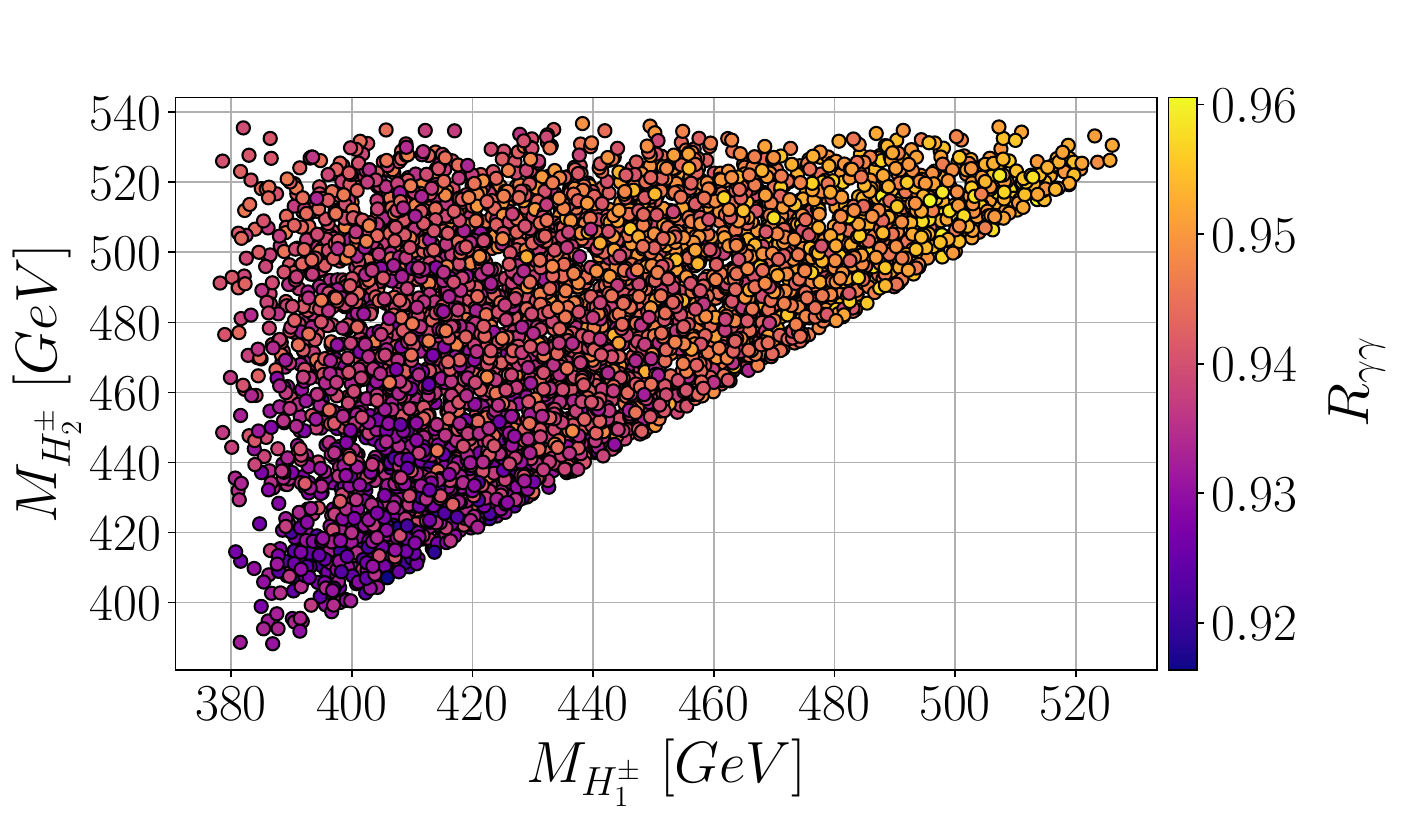}
\caption{Samples allowed in the plane $M_{H_1^\pm}-M_{H_2^\pm}$ in relation to the Higgs di-photon signal strength.}\label{correlation:Rgamma1}
\end{figure}

\begin{figure}[]
\centering
\includegraphics[scale=.6]{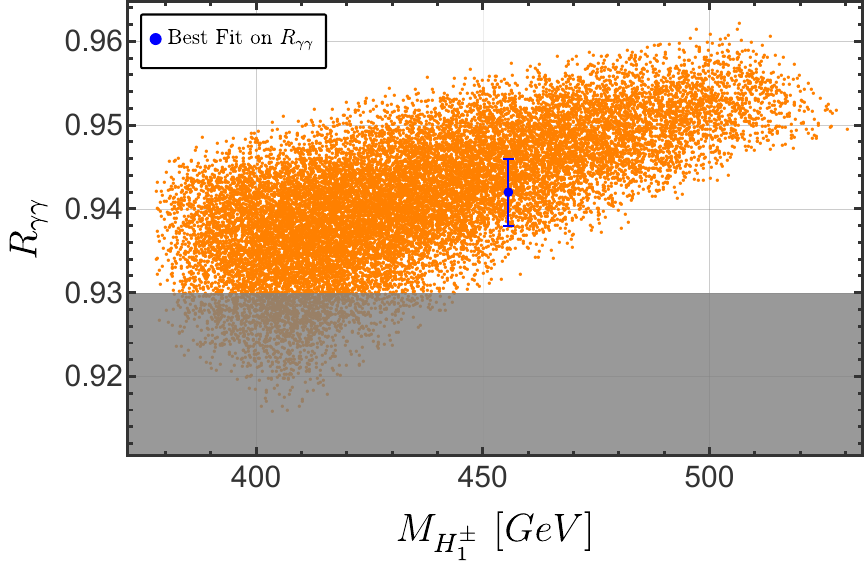}
\caption{The figure illustrates the correlation between the Higgs di-photon signal strength and the mass of the charged scalar $M_{H_1^{\pm}}$. The blue dot represents the best fit for the ratio $R_{\gamma \gamma}$ along with its corresponding uncertainty. The solid region in the plot represents values outside the experimental range at a $1\sigma$ level, indicating inconsistency with the observed data.}
\label{correlation:Rgamma2}
\end{figure}

For the numerical case, considering the ratio $R_{\gamma \gamma}$ has been measured by CMS \cite{Saha:2022cnz} and ATLAS  \cite{ATLAS:2022tnm} collaborations with the best fit signals:
\begin{equation}
R_{\gamma \gamma}^{(\text{CMS})}=1.02_{-0.09}^{+0.11} \quad \text { and } \quad R_{\gamma \gamma}^{(\text{ATLAS})}=1.04_{-0.09}^{+0.10},
\end{equation}
and the best fit result for the ratio $R_{\gamma \gamma}$ is: 
\begin{equation}
R_{\gamma \gamma}^{(1+2\text{HDM})} = 0.942 \pm 0.004.
\end{equation}
Numerically, the central value obtained exhibits a relative error of $0.09$ and $0.08$ concerning the ATLAS and CMS data, respectively, falling within the $1\sigma$ experimentally allowed range. Additionally, Table \ref{table:diphoton} presents the results of the lighter additional scalar masses and the adimensional trilinear couplings between the SM Higgs boson ($h^0$) and the $H_1^{\pm}$ and $H_2^{\pm}$ fields. The similarity between the two values arises from the quasi-degeneracy of the charged scalar fields. Figure~ \ref{correlation:Rgamma1} displays the allowed parameter space of charged scalar masses consistent with the allowed experimental ranges of the Higgs di-photon signal strength. It is evident from Figure~\ref{correlation:Rgamma2} that our model favors a Higgs decay rate into two photons lower than the expectation of the Standard Model but within the allowed range, indicated by the points outside the solid gray region. The blue dot represents the best fit in $R_{\gamma \gamma}^{(1+2\text{HDM})}$, including the uncertainty of $\pm 0.004$ obtained from the sampled range of the charged scalar mass $487.99 \ [\text{GeV}]\leq M_{H_1^{\pm}} \leq 517.79 \ [\text{GeV}]$.

\begin{table}[]
\centering
\begin{tabular}{lc}
\toprule[0.2mm]
\hline  Parameters & Model value \\
\hline \hline$M_{h^0}$ & $125.2 \pm 0.2\ \mathrm{GeV}$ \\
$M_{H_1^0}$ & $ 503 \pm 15\ \mathrm{GeV}$ \\
$M_{A_1^0}$ & $ 501 \pm 15\ \mathrm{GeV}$ \\
$M_{H_1^{\pm}}$ & $455 \pm 16\  \mathrm{GeV}$ \\
$\mathcal{A}_{h H_1^{\pm} H_1^{\mp}}$ & $0.28 \pm 0.03$ \\
$\mathcal{A}_{h H_2^{\pm} H_2^{\mp}}$ &$ 0.28 \pm 0.02$ \vspace{.1cm}\\ 
\hline
\bottomrule[0.2mm]
\end{tabular}
\caption{Parameters with $v=246$ GeV.}
\label{table:diphoton}
\end{table}

\section{The oblique parameters $S$, $T$, $U$}\label{oblique}

The oblique parameters $S$, $T$, $U$ quantify the corrections to the two-point functions of gauge bosons through loop diagrams. 
The extra scalars affect the oblique corrections of the
SM, and these values are measured in high precision
experiments. Consequently, they act as a further constraint
on the validity of our model. The oblique corrections are
parametrized in terms of the  well-known quantities  $S$, $T$, $U$. In this section we calculate one-loop contributions to
the oblique parameters  $S$, $T$, and $U$ defined as \cite{Peskin:1991sw,Altarelli:1990zd,Barbieri:2004qk}

\begin{eqnarray}
 S & = &-\left.\frac{4  c_W s_W}{ \alpha_{\text{em}}} \frac{d}{d q^2} \Pi_{30}\left(q^2\right)\right|_{q^2=0},\\
 T &=& \frac{1}{\alpha_{\text{em}} M_W} \left. \left[ \Pi_{11}\left(q\right) - \Pi_{33} \left( q^2\right)\right] \right|_{q^2=0} \\
 U &= & \frac{4 s_W }{\alpha_{\text{em}} } \left. \frac{d}{d q^2}\left[\Pi_{11}\left(q^2\right)-\Pi_{33}\left(q^2\right)\right]\right|_{q^2=0} ,
 \end{eqnarray}
with $s_W=\sin \theta_W$ and $c_W=\cos\theta_W$, where $\theta_W$ is the electroweak mixing angle, the quantity $\Pi_{ij}(q^2)$ are the vacuum polarization
amplitudes, with $i,j=0,1,3$ for $B$, $W_1$ and $W_3$ gauge bosons respectively. Further details can be found in Ref. \cite{Novikov:1992rj,Peskin:1991sw}.

In the standard model, the oblique parameters are contributed by the SM-Higgs masses and the $W$ boson mass. The contribution from the new physics comes from the extra scalar fields, as noted above. Currently, in the literature, we find the calculations for $2$HDM and $3$HDM in Reds. \cite{CarcamoHernandez:2015smi,CarcamoHernandez:2015mkh}. For our $3$HDM model, we will use the expressions derived in \cite{CarcamoHernandez:2023dyz,Grimus:2007if,Grimus:2008nb}, which represent a generalization for any type of $3$HDM model:
	
\begin{eqnarray} 
\label{Seq} S & \simeq & \frac{1}{12 \pi} \sum_{i=1}^2 \sum_{j=1}^2 \sum_{k=1}^2 \left[\left(R_\alpha\right)_{k i}\left(R_{\gamma}\right)_{k j}\right]^2 K\left(M_{H_i^0}, M_{A_j^0}, M_{H_k^{ \pm}}\right) \\
\nonumber &&\\
\label{Teq} T & \simeq & t_0 \left[ 
\sum_{a=1}^{2}\sum_{k=1}^2\left[\left(R_{\beta}\right)_{a k}\right]^2 M_{H_k^{ \pm}}^2
+\sum_{a=1}^{2} \sum_{i=1}^2 \sum_{j=1}^2\left[ \left(R_\alpha\right)_{a i} \left(R_{\gamma}\right)_{a j}\right]^2 F\left(M_{H_i^0}, M_{A_j^0}\right)
\right. \\
\nonumber & & \left.- \sum_{a=1}^{2}\sum_{i=1}^2 \sum_{k=1}^2 \left\lbrace  \left[\left(R_\alpha\right)_{a i}\left(R_{\beta}\right)_{a k}\right]^2 F\left(M_{H_i^0}, M_{H_k^{\pm}}\right)+
\left[\left(R_{\gamma}\right)_{a i}\left(R_{\beta}\right)_{a k}\right]^2 F\left(M_{A_i^0}, M_{H_k^{ \pm}}\right)\right\rbrace \right]  \\
\nonumber &&\\
\label{Ueq} U &\simeq & -S+\sum_{a=1}^{2}\sum_{i=1}^2 \sum_{k=1}^2  \left\lbrace \left[ \left(R_{\gamma}\right)_{a i}\left(R_{\beta}\right)_{a k}\right]^2 G\left(M_{A_i^0}, M_{H_k^{\pm}}\right)  +\left[\left(R_\alpha\right)_{a i}\left(R_{\beta}\right)_{a k}\right]^2 G\left(M_{H_i^0}, M_{H_k^{\pm}}\right) \right\rbrace 
\end{eqnarray}

where $t_0=\left( \pi^2 v^2 \alpha_{E M}\left(M_Z\right)\right)^{-1}$.
Furthermore, the following loop functions were introduced in \cite{CarcamoHernandez:2015smi,CarcamoHernandez:2015mkh}
\begin{eqnarray}
F\left(z_1,z_2\right)&= & \frac{z_1^2 z_2^2}{z_1^2-z_2^2} \ln \left(\frac{z_1^2}{z_2^2}\right)\\
G\left(z_1, z_2\right) &= & \frac{-5 z_1^6+27 z_1^4 z_2^2-27 z_1^2 z_2^4+6\left(z_1^6-3 z_1^4 z_2^2\right) \ln \left(\frac{z_1^2}{z_2^2}\right)+5 z_2^6}{6\left(z_1^2-z_2^2\right)^3}\\
K\left(z_1, z_2, z_3\right) & = & \frac{1}{\left(z_2^2-z_1^2\right)^3}\left\{z_1^4\left(3 z_2^2-z_1^2\right) \ln \left(\frac{z_1^2}{z_3^2}\right)-z_2^4\left(3 z_1^2-z_2^2\right) \ln \left(\frac{z_2^2}{z_3^2}\right)\right. \nonumber \\
&& \left.-\frac{1}{6}\left[27 z_1^2 z_2^2\left(z_1^2-z_2^2\right)+5\left(z_2^6-z_1^6\right)\right]\right\},
\end{eqnarray}

The inclusion of the two inert doublets, denoted as $H_1$ and $H_2$, in a $1+2$HDM model implies that the oblique parameters $S$, $T$, and $U$ introduce crucial loop 
level corrections to the vacuum polarization amplitudes involving SM gauge bosons in their external lines. Such corrections arise from 
the 
couplings of the scalar fields included in the inert doublets with the Standard Model gauge bosons. 
In equations (\ref{Seq}-\ref{Ueq}),
the symbols $R_{\alpha}$, 
$R_{\beta}$, and $R_{\gamma}$ represent the mixing matrices as presented in Section \ref{scalarsector}, and $M_{H_k^{\pm}}$, $M_{H_k^{0}}$ and $M_{A_k^0}$ 
are the masses of the physical charged, neutral CP even and neutral CP odd dark scalars, respectively.

\begin{figure}{}
\subfigure[]{\includegraphics[scale=.6]{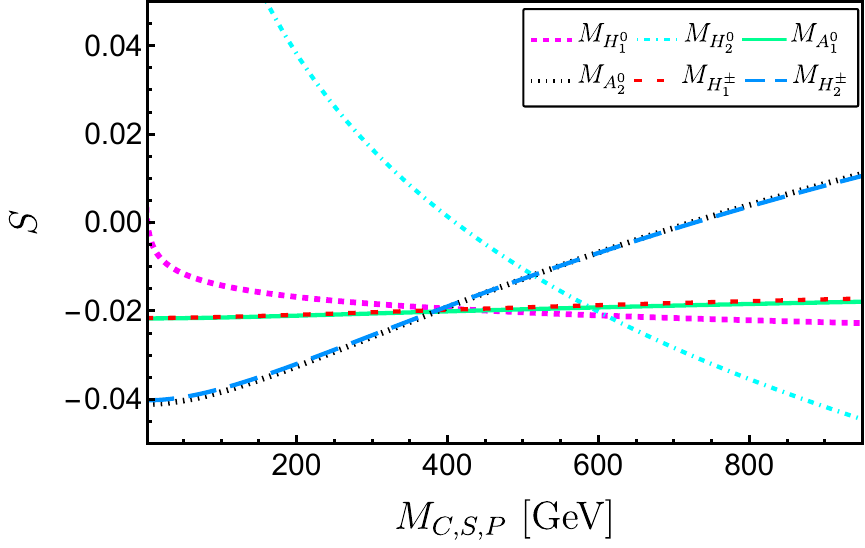}} \quad
\subfigure[]{\includegraphics[scale=.6]{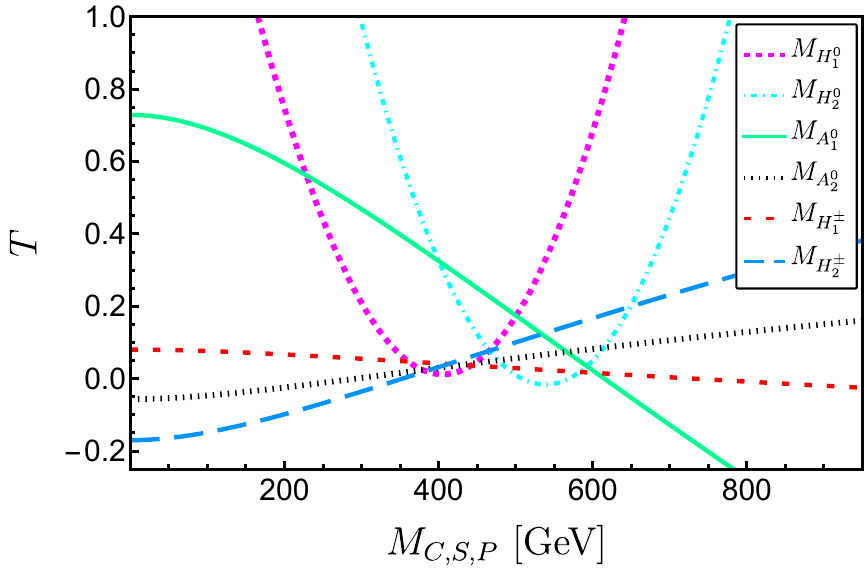}}
\subfigure[]{\includegraphics[scale=.6]{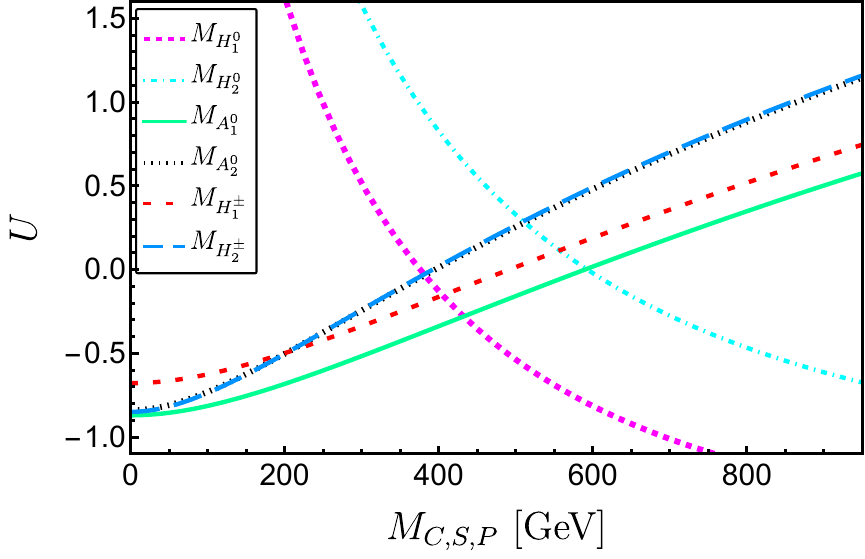}}
\caption{Graphs of the parameters $T$, $S$, and $U$ for the various contributions of the six additional scalar fields in the 3HDM model.}
\label{plotlineSTU}
\end{figure}

To more effectively illustrate these contributions and comprehend their impact on the model, we have generated graphs show in Figure~\ref{plotlineSTU}, representing the dependence of the masses of the charged scalars ($M_C$), neutral scalars ($M_S$), and pseudo-scalar neutral scalars ($M_P$) on the oblique parameters $S$, $T$, and $U$. These graphs are constructed based on previously derived theoretical expressions, providing a visual depiction of how the parameter values vary as the masses of the extra scalars are modified within the range of interest, namely, $100 \leq M_{C,S,P} \leq 900$ GeV.

The obtained best fit point values for the oblique $S$, $T$ and $U$ parameters in our model are: 
\begin{eqnarray}
S &= & -0.019 \pm 0.003,\\
T &= & 0.029\pm 0.007, \\
U &= & 0.010\pm 0.003.
\end{eqnarray}

Our analysis shows consistency with the current experimental limits given in Ref.~\cite{Workman:2022ynf}: for $S_{\text{exp}}  = -0.02 \pm 0.1$, $T_{\text{exp}} = 0.03 \pm 0.12$, and $U_{\text{exp}}  = 0.01 \pm 0.11$.

\begin{figure}[]
\centering
\subfigure[]{\includegraphics[scale=.25]{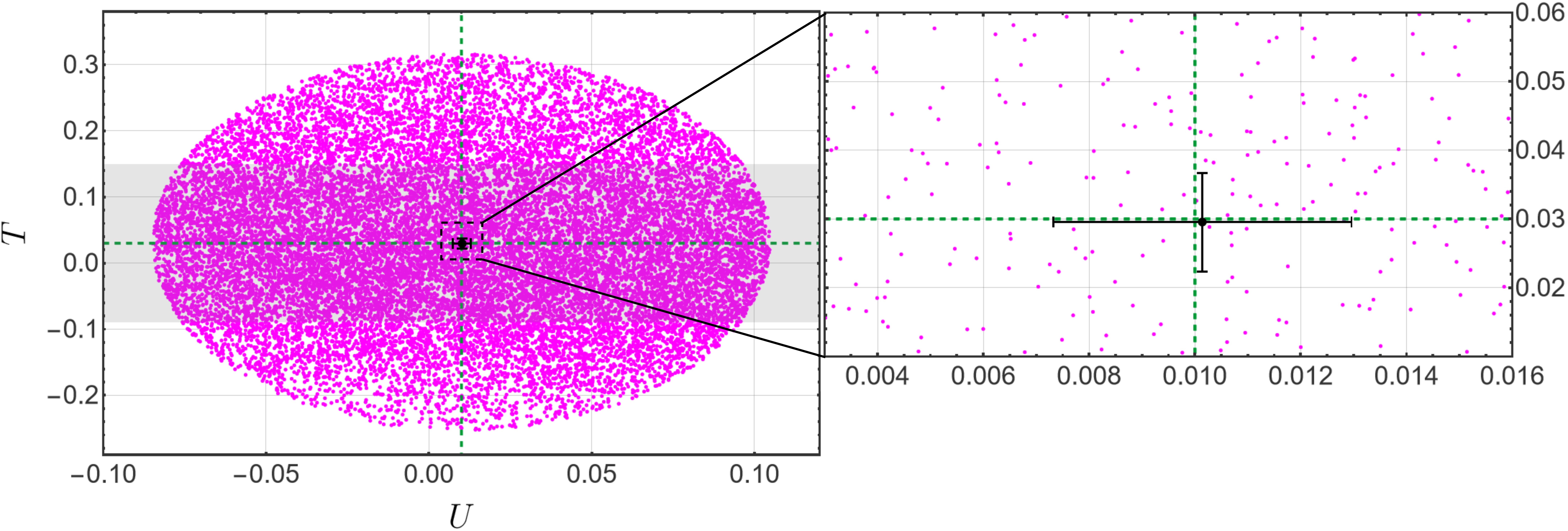}}
\subfigure[]{\includegraphics[scale=.42]{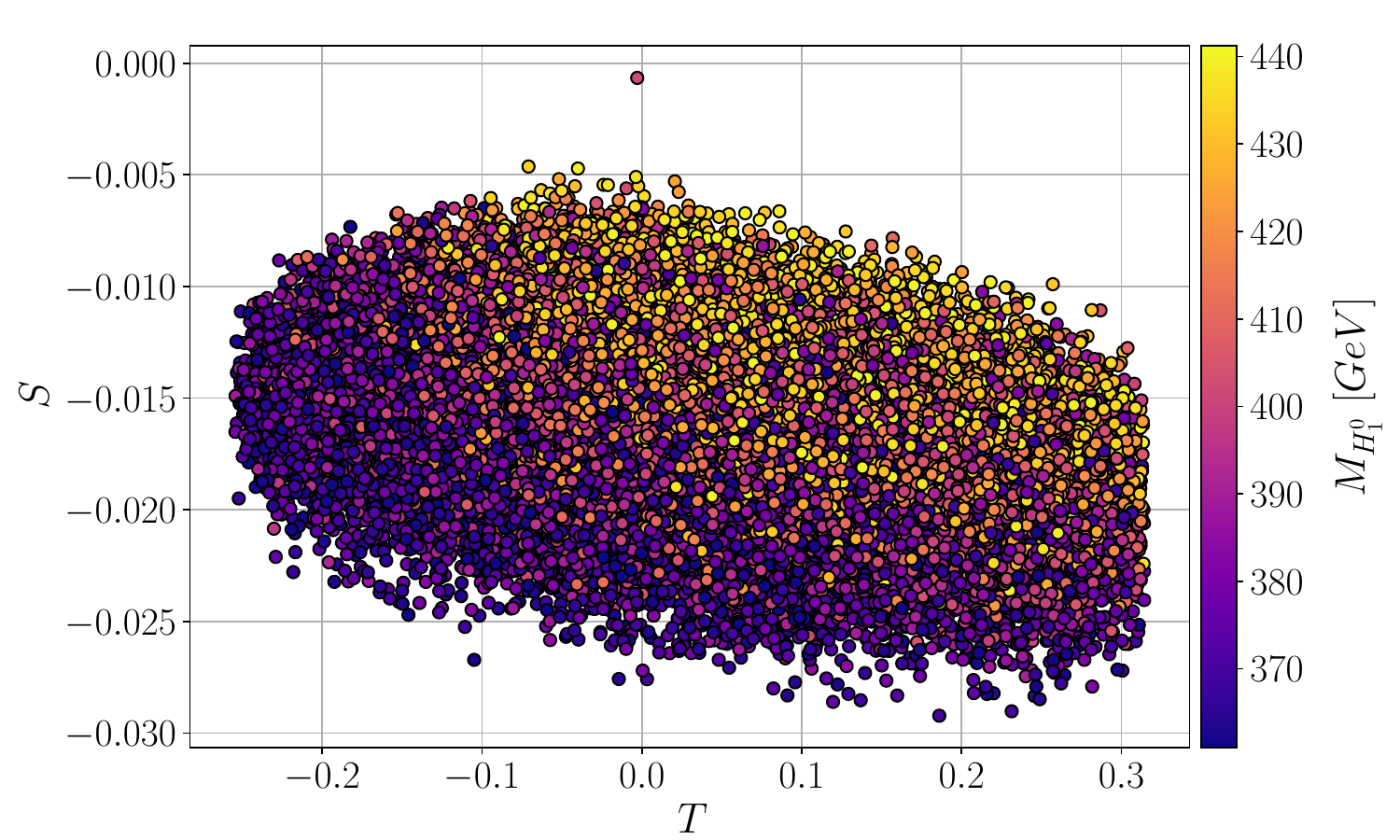}}
\subfigure[]{\includegraphics[scale=.37]{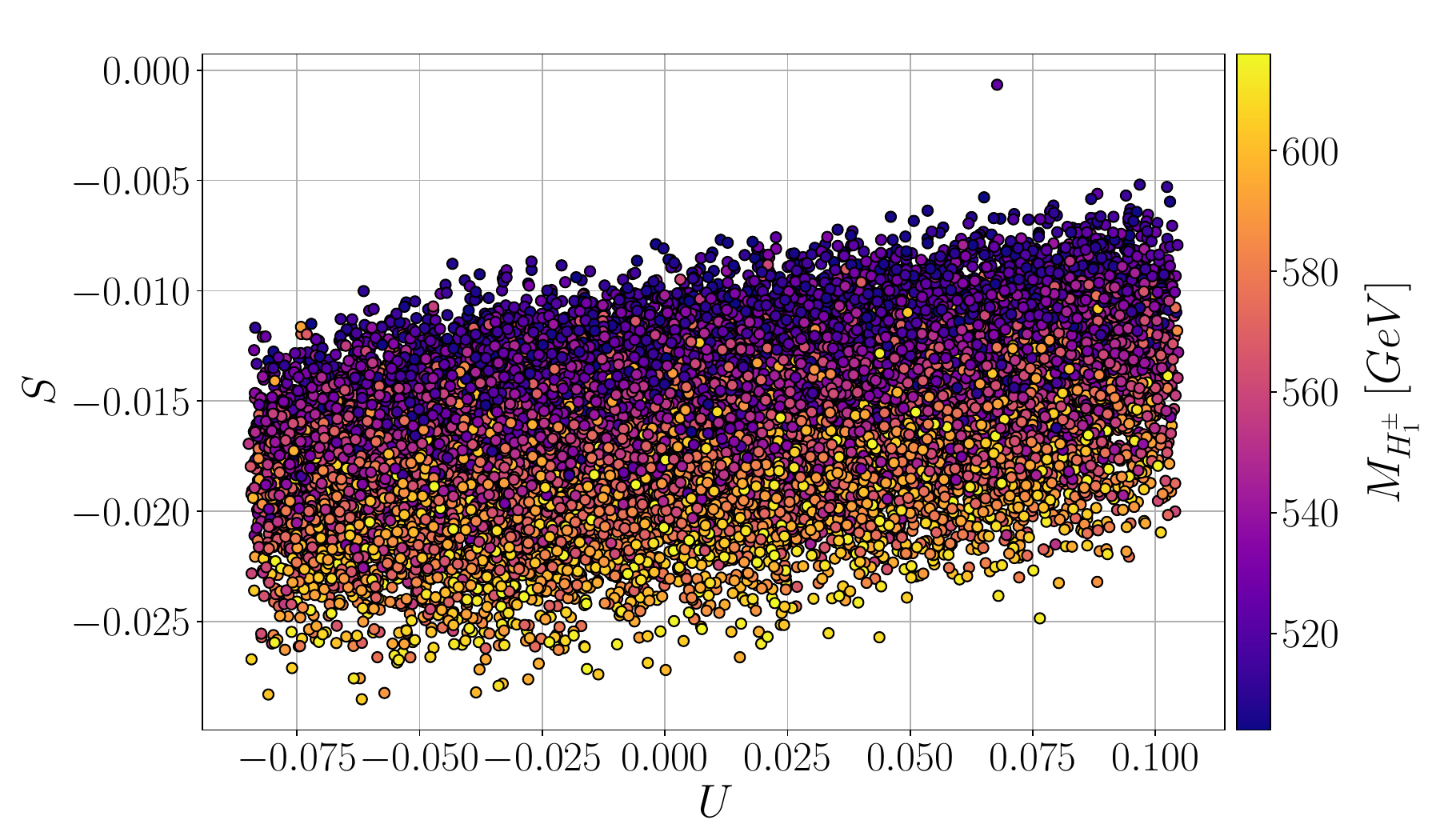}}
\caption{(a) Correlation between the oblique parameters $U$ and $T$. The green vertical and horizontal segmented line indicates the central experimental values of $U$ and $T$ respectively \cite{Workman:2022ynf}, while the gray region points to allowed values at $1 \sigma$ of the $T$ parameter. The black dot indicates the best-fit point together with its respective uncertainty. (b) Correlation between the oblique parameters $T$ and $S$. The color represents the scan of the mass $M_{H_1^{0}}$ as a function of the parameters. (c) Correlation between the oblique parameters $U$ and $S$. The color represents the scan of the mass $M_{H_1^{0}}$ as a function of the parameters.}
\label{fig:obliquecorrelation}
\end{figure}

Figure~\ref{fig:obliquecorrelation} displays the correlation between the oblique $S$, $T$ and $U$ parameters and indicates
that the values are consistent with the constraints derived from the experimental measurements. In (a) the dashed lines are the central experimental values, the grey region corresponds to the experimental value at $1 \sigma$ of $T$, in the enlarged figure the black dot represents the value obtained by the model. In (b) the colour represents the dispersion with respect to the mass of the particle $H_1^0$, the scattering of the points shows a certain symmetry with respect to $T=0$, where colder colours like violet and blue dominate, for extreme values of $T$ are associated with higher mass values, and warmer colours like yellow and red take positive and negative values. In (c) the colour represents the dispersion with respect to the mass of $H_1^{\pm}$, for high values of the mass represented by warm colours are grouped in the lower part of the graph favoring $S \sim -0.025$, while in the upper region cooler colours (smaller masses) are grouped $S \sim -0.07$, at $U=0$ there is symmetry in the region.

\section{Scalar Dark Matter}\label{DM}

In this section we analyze the implications of the model in dark matter. Due to the preserved $Z_2$ symmetry, our model has dark matter candidate which will be the lightest of the electrically neutral particles among having non trivial $Z_2$ charges. As indicated in Table
\ref{leptons} the $SU(2)$ scalar doublets $H_1$ and $H_2$ as well as the right handed Majorana neutrinos are $Z_2$ odd. Therefore, the dark matter candidate in our model will corresponds to the lightest among the $H_1^{0}$, $H_2^{0}$, $A_1^{0}$, $A_2^{0}$ and $N_R$ fields. In our analysis of dark matter we choose a benchmark scenario where the right handed Majorana neutrino $N_R$ is heavier than the dark scalars $H_1^{0}$, $H_2^{0}$, $A_1^{0}$, $A_2^{0}$, then implying that the dark matter candidate in our model is a scalar.
\begin{figure}[H]
\centering
\includegraphics[scale=.45]{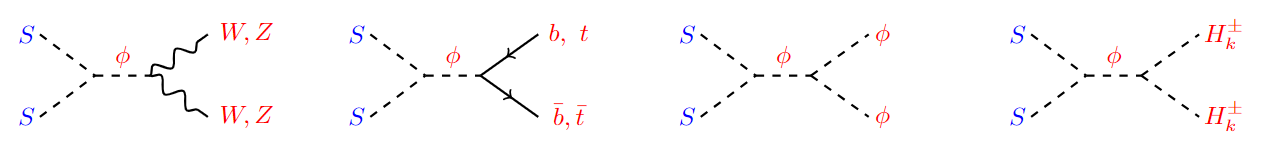}
\caption{The dark matter annihilation to standard model and extra charged scalar particles for the
benchmark considered. $S$ indica el candidato escalar dark matter.}
\label{fig:feymandiagramDM}
\end{figure}
In the model under consideration, the scalar dark matter (DM) candidate, corresponding to the lightest among the $H_1^0$, $H_2^0$, $A_1^0$, and $A_2^0$ fields, interact with Standard Model particles only through a Higgs portal scalar interaction. This is illustrated in Figure~\ref{fig:feymandiagramDM}, which shows the $2 \rightarrow 2$ annihilation processes of a dark matter candidate $S$ into Standard Model particles governed by the quartic scalar coupling $\lambda_{\text{DM}}$.
Considering $A_1^0$ as the lightest candidate for dark matter, it is reasonable to assume that the dark matter annihilation process occurs predominantly in the $s$-channel, due to the operators $\phi^{\dagger} \phi H_{1}^{\dagger} H_{1}$ and $\phi^{\dagger} \phi H_{2}^{\dagger} H_{2}$. In contrast, the channel-$t$ processes, which would involve the exchange of a dark matter particle with a neutrino, are highly suppressed due to the small effective neutrino Yukawa couplings. This suppression makes the channel-$s$ processes dominant in dark matter annihilation in this model.

Thus, we can determine the reaction rate by considering distributions at thermal equilibrium and adjusting them to accommodate the measured value of the 
current dark matter relic density. The cross-sections of DM annihilations into SM particles are given by \cite{Hambye:2009pw,Bhattacharya:2016ysw}

\begin{equation}
\begin{aligned}
\langle\sigma v\rangle_{S S \rightarrow V V}= & \ \frac{\lambda_{\text{DM}}^2}{ 8 \pi\kappa_V} \frac{s}{\left(s-m_h^2\right)^2+m_h^2 \Gamma_h^2}\left(1+\frac{12 m_V^4}{s^2}-\frac{4 m_V^2}{s}\right) \sqrt{1-\frac{4 m_V^2}{s}}, \\
\langle\sigma v\rangle_{S S \rightarrow q \bar{q}}= & \ \frac{\lambda_{\text{DM}}^2}{4 \pi s \sqrt{s}} \frac{N_c  m_q^2}{\left(s-m_h^2\right)^2+m_h^2 \Gamma_h^2}\left(s-4 m_q^2\right)^{\frac{3}{2}} \\
\langle\sigma v\rangle_{S S \rightarrow h h}= & \ \frac{\lambda_{\text{DM}}^2}{16 \pi s}\left[1+\frac{3 m_h^2}{\left(s-m_h^2\right)}-\frac{4 \lambda_{\text{DM}} v^2}{\left(s-2 m_h^2\right)}\right]^2  \sqrt{1-\frac{4 m_h^2}{s}}
\end{aligned}
\label{DMannihilationsSM}
\end{equation}
where $\sqrt{s}$ is the squared center-of-mass energy, $\Gamma_h$ denotes the total SM Higgs decay width 
which is equal to 
$4.1$ MeV,  and $\kappa_V$ is a factor equal to $1$ and $2$ for the $W^{\pm}$ and $Z$ bosons respectively. The benchmark also considers DM annihilation processes into other extra charged scalar particles in the $s$ channel,
controlled by the portal scalar coupling $\lambda_{\text{E}}$ into a pair of non-SM electrically charged scalars. The corresponding annihilation cross-section into a pair of non SM electrically charged scalars can be written as follows:
\begin{equation}
\langle\sigma v\rangle_{S S \rightarrow H_{k}^{\pm} H_{k}^{\pm}}=\frac{\lambda_{\text{E}}^2}{16 \pi s} \sqrt{1-\frac{4 m_{H^{\pm}_{k}}^2}{s}}
\label{DMannihilationsNONSM}
\end{equation}
where the final state represented as $H_{k}^{\pm}$ ($k=1,2$) stands for the electrically charged scalars. 

\begin{figure}[]
\centering
\subfigure[]{\includegraphics[scale=.6]{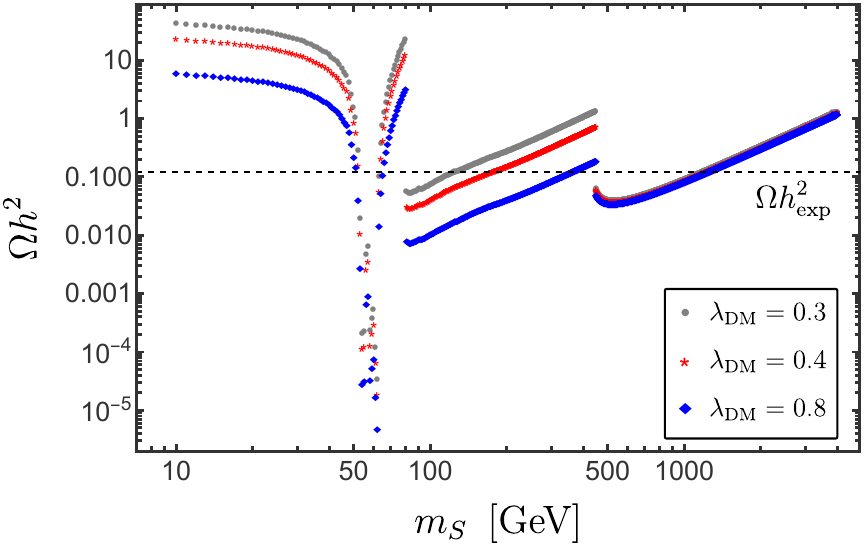}}\quad 
\subfigure[]{\includegraphics[scale=.6]{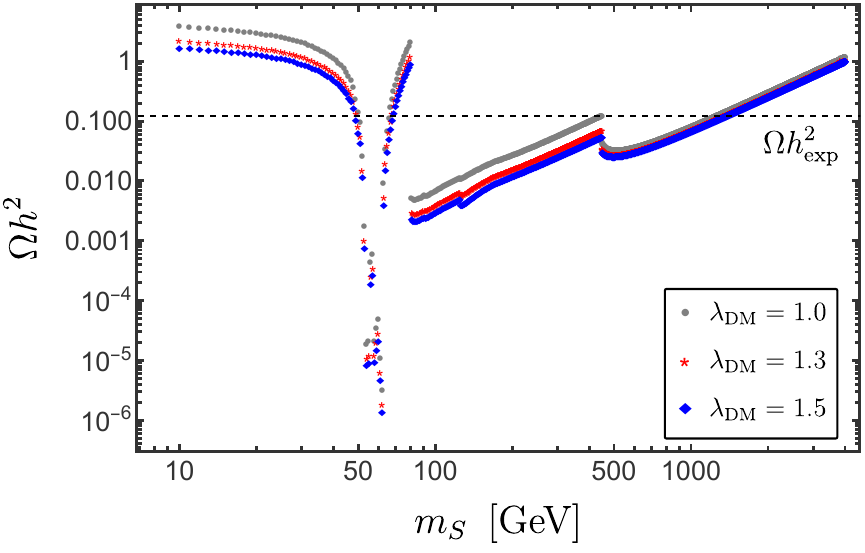}}
\caption{Dark matter relic density  $\Omega h^2$ as a function of the dark matter mass  $m_S$ for different values of the portal coupling $\lambda_\text{DM}$ (a) $\{ 0.3,0.4,0.8\}$ and (b) $\{ 1.0,1.3,1.5\}$ for gray circle, red star and blue rhombus respectively. The black dashed line corresponds to central value $\Omega h^2_{\text{exp}}=0.1200$ \cite{Planck:2018vyg}.}
\label{fig:relicdensity}
\end{figure}

The dark matter relic abundance in the present Universe is estimated as follows
\begin{equation}
\Omega h^2=\frac{0.1 \mathrm{pb}}{\langle\sigma v\rangle}, \quad \quad \langle\sigma v\rangle=\frac{A}{n_{e q}^2},
\end{equation}
where $\langle\sigma v\rangle$ is the thermally averaged annihilation cross section, and $A$ is the total annihilation rate per unit volume at temperature $T$ and $n_{eq}$ is the equilibrium value of the particle
density, which are given as
\begin{equation}
\begin{aligned}
A & =\frac{T}{32 \pi^4} \int_{4 m_s^2}^{\infty} \sum_{X} g_X^2  \langle\sigma v\rangle_{S S \rightarrow X X} \frac{s \sqrt{s-4 m_\varphi^2}}{2} K_1\left(\frac{\sqrt{s}}{T}\right) d s, \\
n_{e q} & =\frac{T}{2 \pi^2} \sum_{X} g_X m_\varphi^2 K_2\left(\frac{m_\varphi}{T}\right)
\end{aligned}
\end{equation}
with $K_1$ and $K_2$ being the modified Bessel functions of the second kind of order 1 and 2, respectively, here $X$ correspond $t$, $b$, $Z$, $W^\pm$, $\phi$, $H_k^{\pm}$. We have taken a freeze-out temperature $T=m_{S} / 20$ following Ref.\cite{Edsjo:1997bg}. Figure~\ref{fig:relicdensity} displays the dark matter relic density as a function of the scalar dark matter mass $m_S$. Here we have used Eqs. (\ref{DMannihilationsSM}-\ref{DMannihilationsNONSM}) in the DM annihilation in $W^{\pm}W^{\pm}$, $ZZ$, $\phi \phi$, $\bar{t}t$, $\bar{b}b$, and the charged components of the inert scalar doublets, we also show a dashed horizontal line the value measured experimentally by Planck 2018 results for the relic density with a value \cite{Planck:2018vyg}
\begin{equation}
\Omega h^2_{\text{exp}}=0.1200 \pm 0.0012.
\end{equation}
In Figure~\ref{fig:relicdensity}, three different mass regions are analyzed: low mass $m_S < 50$ GeV, intermediate mass $60 < m_S <300$ GeV, and high mass $m_S>400$ GeV. These demarcations are due to thresholds in the dark matter (DM) annihilation processes. The first gap occurs at $M_{h}/2$, where the annihilation channel is highly effective due to the on-shell nature of the Standard Model (SM) Higgs. The second gap corresponds to the $ZZ$ annihilation process near the $Z$-boson mass, $M_Z$. The last discontinuity emerges around $m_S \simeq M_{A_k^0}$, where the DM annihilation channel to new particles proposed by the model opens. These findings are consistent with results from micrOMEGAs \cite{Belanger:2018ccd}. The most important contribution to Dark Matter annihilation arises from the channel that produces a pair of $Z$ bosons with $41$\% then then a pair of tau leptons with 16 \%, a pair of $W^{\pm}$ leptons with $14$\%. In the plot of the left panel for the low mass region $m_S < 50$ GeV, the DM relic abundance would exceed the observed levels, indicating that for these masses, the DM candidate is not viable or requires additional mechanisms. When the mass is increased, small regions of underabundance are noted; however, they alone cannot explain all DM. In the right plot, where the portal coupling is larger, a similar behavior is observed for masses below $50$ GeV. The threshold at half the mass of the SM Higgs is higher than that for portal couplings less than unity, significantly favoring the DM annihilation process. For $m_S>70$ GeV, there is a better alignment for predicting a WIMP-type DM candidate unless other factors or types of dark matter are considered. 

\begin{figure}[]
\centering
\includegraphics[scale=.7]{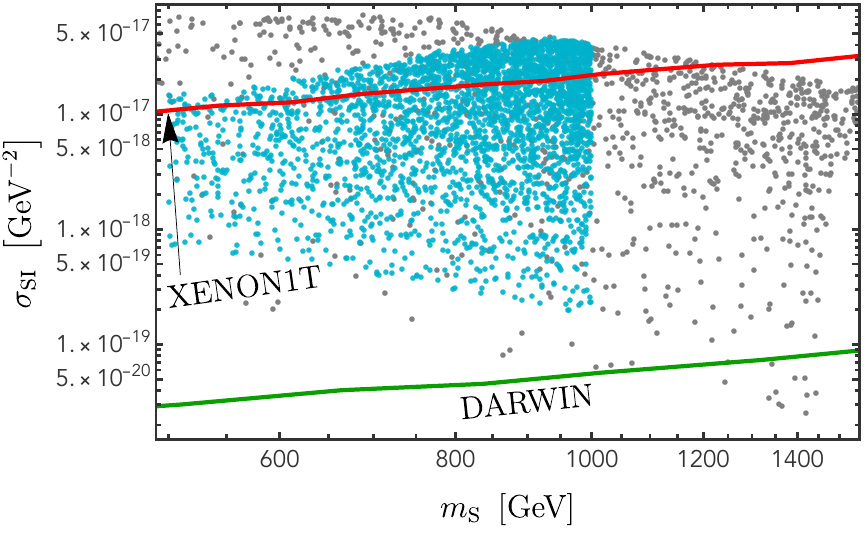}
\caption{Spin independent direct detection cross section $\sigma_{\text{SI}}$ as a function of dark matter mass. 
The points show the values of the DM relic density consistent with the experimental constraints. The solid red line denotes the bound arising from the XENON1T experiment \cite {XENON:2018voc} while the green line represents the projected sensitivities of DARWIN \cite{DARWIN:2016hyl}. The cyan and gray dots favour smaller and larger coupling portal values respectively.}
\label{fig:sigmaDD}
\end{figure}

Direct detection is another important constraint to consider when studying the feasibility of the model. The elastic scattering of dark matter particles with nuclei is made possible by Higgs portal interactions, driven by $\lambda_{\text{DM}}$. The expression for the spin-independent effective cross section (SI) is formulated as follows \cite{Bernal:2017xat}:
\begin{equation}
\sigma_{\mathrm{SI}}=\frac{\lambda_{\text{DM}}^2 m_N^4 f_N^2}{8 \pi M_h^4 m_{S}^2}
\end{equation}
where $m_N$ is the mass of the nucleon for direct detection. Here, the coupling $f_N$ is the matrix element that depends on the quark content in the nucleon for each flavor of quark and is expressed as
\begin{equation}
f_N=\sum_q f_q=
 \sum_q \frac{m_q}{m_N} \int d^3r \, \phi_q^\dagger(r) \phi_q(r)
\end{equation}
where the sum is over all quark flavours, $\phi_q(r)$ is the quark wave function in the nucleon and $m_q$ is the quark mass. The heavy quark contributions are expressed in terms of the light quarks via specific theoretical relations in Ref. \cite{Shifman:1978zn}. In Figure~\ref{fig:sigmaDD} we present a DM prediction of direct spin-independent detection as a function of the DM candidate consistent with the relic density $\Omega h^2$, taking $f_N=1/3$ \cite{Cirelli:2005uq}. In this model, the parameter scan shows cyan points corresponding to lower values of portal coupling, while the gray points correspond to a larger sweep of portal coupling favouring values above 1. In both cases, there are favourable points for direct detection below the red line bounded by XENON1T \cite{XENON:2018voc} and above the green DARWIN experimental line \cite{DARWIN:2016hyl}, which delimits an upper region for our $1+2$HDM model.

\section{Conclusions}\label{conclusion}
We have constructed a scalar singlet extension of the 1+2 Higgs doublet model, where the leptonic mixings is governed by a perturbed cobimaximal mixing pattern. The theory under consideration, whose scalar spectrum contains one active $SU(2)$ scalar doublet, identified as the SM Higgs doublet, two inert $SU(2)$ scalar doublets as well as some singlet scalars, is based on the $S_3$ family symmetry supplemented by the preserved $Z_{2}$ symmetry and the spontaneously broken $S_{3}\otimes Z_{2}^{\prime} \otimes Z_{18}$ discrete group, allows for a successfull fit of the charged lepton masses, neutrino mass squared differences, leptonic mixing angles and leptonic Dirac CP phase. The active neutrino masses are generated from a radiative seesaw mechanism at one loop level and the observed SM charged lepton mass hierarchy is originated from the spontaneous breaking of the $S_{3}\otimes Z_{2}^{\prime} \otimes Z_{18}$ discrete group. The theory has stable scalar and fermionic dark matter candidates whose stability is ensured by the preserved $Z_2$ symmetry, which also guarantee the radiative nature of the seesaw mechanism responsible for the generation of the active neutrino masses. We performed a comprehensive analysis of the scalar sector of the theory as well as of its implications in oblique paramaters, Higgs diphoton decay, charged lepton flavor violation and dark matter. We found that the model under consideration is consistent with the constraints arising from tree level stability of the scalar potencial, oblique $S$, $T$ and $U$ parameters, SM Higgs decay into two photons, charged lepton flavor violating decays and dark matter. Further, our model yields rates for charged lepton flavor violtating decays $\mu \rightarrow e\gamma$, $\mu \rightarrow 3e$ as well as the muon-electron conversion processes, within the reach of sensitivity of experiments. Finally, we analyzed in detail the scenario of scalar dark matter candidate finding that our model successfully complies with the constraints arising from dark matter relic density and dark matter direct detection. 

\section*{Acknowledgments}

This work as received funding from ANID PIA/APOYO AFB230003, Fondecyt 1230110 (Chile), Fondecyt 1210131 (Chile), and ANID-Chile Fondecyt 1210378, ANID- Programa Milenio - code ICN2019\_044.

\appendix

\section{The product rules for $S_{3}$}\label{appen:s3}

\begin{figure}
\centering
\includegraphics[scale=0.5]{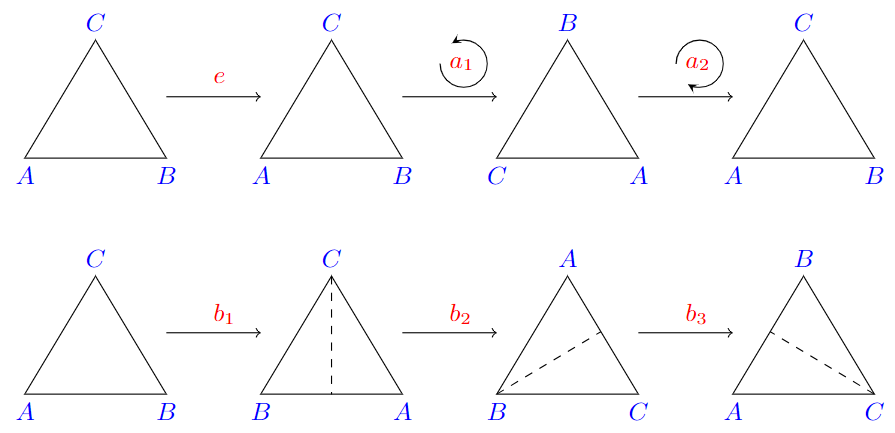}    
\caption{The $S_3$ symmetry (the rotations and the reflections) of an equilateral triangle.}
\label{fig:S3}
\end{figure}

The group $S_3$ is the permutation group of three objects, which can be geometrically represented by the different rotations that leave invariant an equilateral triangle and has six elements, the smallest number of elements in non-Abelian discrete groups, which we represent by $e$, $a_1$, $a_2$, $b_1$, $b_2$, $b_3$. The elements $e$, $a_1$, $a_2$ are cyclic permutations, $b_1$, $b_2$, $b_3$ are anti-cyclic, the element $e$ corresponds to the identity, as shown in Figure \ref{fig:S3}. The multiplication table is given below \cite{Deshpande:1991zh}:
\begin{equation}
\begin{array}{c|cccccc}
S_3 & e & a_1 & a_2 & b_1 & b_2 & b_3 \\
\hline e & e & a_1 & a_2 & b_1 & b_2 & b_3 \\
a_1 & a_1 & a_2 & e & b_2 & b_3 & b_1 \\
a_2 & a_2 & e & a_1 & b_3 & b_1 & b_2 \\
b_1 & b_1 & b_3 & b_2 & e & a_2 & a_1 \\
b_2 & b_2 & b_1 & b_3 & a_1 & e & a_2 \\
b_3 & b_3 & b_2 & b_1 & a_2 & a_1 & e
\end{array}
\end{equation}
The group contains $3$ irreducible representations, \textit{i.e.} $\mathbf{1}$, $\mathbf{1}^{\prime }$ and $\mathbf{2}$.
Fulfilling the following tensor product rules:
\begin{equation}
    \mathbf{2} \otimes \mathbf{2}=\mathbf{1} \oplus \mathbf{1}^{\prime} \oplus \mathbf{2}, \quad \mathbf{2} \otimes \mathbf{1}^{\prime}=\mathbf{2}, \quad \mathbf{1}^{\prime} \otimes \mathbf{1}^{\prime} = \mathbf{1}.
\end{equation}

Considering $\left( x_1 ,x_2 \right) ^T$\ and $\left( y_1 ,y_2 \right) ^T$
as the basis vectors for two $S_3 $ doublets and $(y{\acute{}})$ is an $S_3 $
non trivial singlet, the multiplication rules of the $S_3 $ group for the
case of real representations take the form \cite{Ishimori:2010au}: 
\begin{eqnarray}
&&\left( 
\begin{array}{c}
x_1 \\ 
x_2%
\end{array}
\right) _{\mathbf{2 }}\otimes \left( 
\begin{array}{c}
y_1 \\ 
y_{2}%
\end{array}%
\right) _{\mathbf{2}}=\left( x_1 y_1 +x_2 y_2 \right) _{\mathbf{1}}+\left(
x_1 y_2 -x_2 y_1 \right) _{\mathbf{1}^{\prime }}+\left( 
\begin{array}{c}
x_2 y_2 -x_1 y_1 \\ 
x_1 y_2 +x_2 y_1%
\end{array}%
\right) _{\mathbf{2}},  \label{6} \\
&&\left( 
\begin{array}{c}
x_1 \\ 
x_2%
\end{array}%
\right) _{\mathbf{2}}\otimes \left( y
{\acute{}}
\right) _{\mathbf{1}^{\prime }}=\left( 
\begin{array}{c}
-x_2 y
{\acute{}}
\\ 
x_1 y
{\acute{}}
\end{array}
\right) _{\mathbf{2}},\hspace{1cm}\left( x{\acute{}}
\right) _{\mathbf{1}^{\prime }}\otimes \left( y
{\acute{}}
\right) _{\mathbf{1}^{\prime }}=\left( x
{\acute{}}
y
{\acute{}}
\right) _{\mathbf{1}}.  \label{7}
\end{eqnarray}

\section{The doublet scalar potential for a $S_3$, decoupling and VEVs} \label{decoupling}

The mixing between the single scalars and doublets of $SU(2)_L$ is suppressed by Eq. \eqref{vevs}, \textit{i.e.}, the scalar singlets of the SM acquire VEVs much larger than the electroweak symmetry breaking scale, so the mixing between single scalars and doublets can be suppressed without loss of generality.

The scalar potential for a $S_3$ doublet $\chi$ is given
\begin{equation}
V_\chi=  - \mu_\chi^2  \left(\chi \chi \right)_{\textbf{1}} 
+\gamma_{\chi}(\chi \chi)_\textbf{2} \chi
+\kappa_{\chi, 1}(\chi \chi)_\textbf{1}(\chi \chi)_\textbf{1} 
 +\kappa_{\chi, 2}(\chi \chi)_\textbf{2}(\chi \chi)_\textbf{2}
 +\kappa_{\chi, 3}\left[(\chi \chi)_\textbf{2} \chi\right]_\textbf{2} \chi,
\end{equation}

From the minimization conditions of the high-energy scalar potential, we find the following relations:

\begin{equation}\label{vev2}
\begin{aligned}
\frac{\partial\langle V_\chi\rangle}{\partial v_{\chi_1}}= & \ - 2 \mu_{\xi}^2 v_{\chi_1} + 3 \gamma_\chi \left(v_{\chi_2}^2 -v_{\chi_1}^2 \right) 
+4 v_{\chi_1} \left(\kappa_{\chi, 1}+\kappa_{\chi, 2}+\kappa_{\chi, 3}\right)\left(v_{\chi_1}^2+v_{\chi_2}^2\right) = 0 \\
\frac{\partial\langle V_\chi\rangle}{\partial v_{\chi_2}}= & \ - 2 \mu_{\xi}^2 v_{\chi_2} + 3 \gamma_\chi v_{\chi_1}v_{\chi_2} 
+4 v_{\chi_2}\left(\kappa_{\chi, 1}+\kappa_{\chi, 2}+\kappa_{\chi, 3}\right)\left(v_{\chi_1}^2+v_{\chi_2}^2\right) = 0 
\end{aligned}
\end{equation}

Then, from an analysis of the minimization equations given by \eqref{vev2}, we obtain for a large range of the parameter space the following VEV direction for $\chi$:
\begin{equation}
 \langle \chi\rangle = \frac{v_{\chi}}{\sqrt{2}}\left(e^{i \theta} , e^{-i \theta}\right).
\end{equation}

Using the vacuum configuration of the above expression and what is found in \eqref{vev2}, we find the relationship between the parameters and the magnitude of the VEV:
\begin{equation}
\mu_{\chi}^2 = 2 v_{\chi}^2 \left( \kappa_{\chi, 1}+\kappa_{\chi, 2}+\kappa_{\chi, 3}\right) \cos\left(2\theta \right)+ \frac{3}{\sqrt{2}}e^{i\theta} \gamma_\chi v_{\chi}.
\end{equation}

On the other hand, the scalar potential $S_3$ is for the doublet $\xi$:
\begin{equation}
V_ \xi=  - \mu_\xi^2  \left( \xi  \xi \right)_{\textbf{1}} 
+\kappa_{ \xi, 1}( \xi  \xi)_\textbf{1}( \xi  \xi)_\textbf{1} 
 +\kappa_{ \xi, 2}( \xi  \xi)_\textbf{2}( \xi  \xi)_\textbf{2}
 +\kappa_{ \xi, 3}\left[( \xi  \xi)_\textbf{2}  \xi\right]_\textbf{2}  \xi,
\end{equation}

Applying the minimization condition of the high-energy scalar potential, we find the following constraint:
\begin{equation}\label{vev3}
\begin{aligned}
  \frac{\partial\langle V_\xi \rangle }{\partial v_{\xi_1}} = & \ - 2\mu_\xi^2 v_{\xi_1}+ 4 v_{\xi_1}\left( \kappa_{ \xi, 1}+\kappa_{ \xi, 2}+\kappa_{ \xi, 3} \right) 
 \left(v_{\xi_1}^2+v_{\xi_2}^2 \right) =0   \\
 \frac{\partial\langle V_\xi \rangle }{\partial v_{\xi_2}} = & \ - 2\mu_\xi^2 v_{\xi_2} + 4 v_{\xi_2}\left( \kappa_{ \xi, 1}+\kappa_{ \xi, 2}+\kappa_{ \xi, 3} \right)  \left(v_{\xi_1}^2+v_{\xi_2}^2 \right) =0
\end{aligned}
\end{equation}

Then, from an analysis of the minimization equations given by \eqref{vev3}, we obtain for a large range of the parameter space the following VEV direction for scalar $\xi$:
\begin{equation}
 \langle \xi\rangle = v_{\xi}\left(1,0\right).
\end{equation}

Using the vacuum configuration de la expresión anterior y lo encontrado en \eqref{vev3}, we find the relation between the parameters and the magnitude of the VEV:
\begin{equation}
\mu_{\xi}^2 = -2 v_{\xi} \left( \kappa_{ \xi, 1}+\kappa_{ \xi, 2}+\kappa_{ \xi, 3} \right).
\end{equation}

These results show that the VEVs direction for the $S_3$ doublet $\chi$ and $\xi$ in \eqref{VEV} is consistent with a global minimum of the scalar potential of our model.

\bibliographystyle{utphys}
\bibliography{ref}

\end{document}